# First-principles predictions of the diversity in atomic structures and electronic properties of the reconstructed Si(111)-7×7 surface

Yuke Song[1,2], ShiFang Li[1,2], PeiZe Lin[3], Jin Li[1,2], Tao Ouyang[1,2], Chao Tang[1,2] and Chaoyu He,[1,2,*]

1. School of Physics and Optoelectronics, Xiangtan University, Hunan 411105, China;
2. Hunan Provincial Key Laboratory of Computational Condensed Matter Physics and Quantum Materials Engineering, Hunan 411105, China;
3. Institute of Artificial Intelligence, Hefei Comprehensive National Science Center, Hefei 230026, Anhui, China;

Electronic address: *hechaoyu@xtu.edu.cn;

The 7×7 reconstruction of Si(111) surface is widely understood by the dimer-adatom-stacking-fault model (DAS), but the predicted metallicity of DAS contradicts experimental signs of insulation. It is still challenge to predict DAS-like reconstructions by traditional method to solve such a puzzle. Here, we show that low-energy reconstructions of Si(111)-7×7 surface with (DAS-$d_8$-$T^{12}$, DAS-$d_8$-$T^9H^{3\text{-}A}$, DAS-$d_8$-$T^9H^{3\text{-}B}$ and DAS-$d_8$-$T^6H^6$) and without (AB-$d_{10}$-$T^{12}$, AB-$d_{10}$-$T^9H^3$, AA-$d_{10}$-$T^{12}$ and AA-$d_{10}$-$T^9H^3$) stacking-fault can be quickly discovered by graph theory as implemented in $RG^2$ code for crystal structure prediction. They exhibit comparable stability to the DAS (DAS-$d_8$-$T^{12}$) model and similar STM patterns, offering a plausible explanation for the observed Si(111)-7×7 reconstruction. All these reconstructions exhibit metallic behavior in the nonmagnetic (NM) state with isolated narrow bands crossing the Fermi level in varying occupancy. And they are further confirmed as ferromagnetic (FM) metals (DAS-$d_8$-$T^9H^{3\text{-}B}$), half-metals (DAS-$d_8$-$T^{12}$, AB-$d_{10}$-$T^9H^3$, AA-$d_{10}$-$T^{12}$ and AA-$d_{10}$-$T^9H^3$), half-semimetals (DAS-$d_8$-$T^9H^{3\text{-}A}$ and DAS-$d_8$-$T^6H^6$) and even insulators (AB-$d_{10}$-$T^{12}$), depending their occupancies of the NM band structures. These findings not only demonstrate the rich electromagnetic phases of reconstructed Si(111) surfaces and their potential for spintronic applications, but also provide a plausible physical explanation for the metal–insulator transition observed on the Si(111) surface.

Crystal surfaces form in the situation of periodicity-broken in one some given direction (hkl). The breaking of symmetry and changes in the coordination environment increase the energy, making the surface energetically or dynamically unstable. Forced by low-energy principles, surface atoms will reconstruct with each other to lower the total energy and form different reconstructions. Such reconstructions can occur even on highly symmetric (111) surfaces in covalent crystals, such as silicon, germanium, silicon carbide etc [1-5]. In the case of silicon, the 7×7 reconstruction of (111) surface has attracted widespread interest in the past decades [4-9] due to its fascinating STM pattern. Extensive experimental efforts have been dedicated to investigating its atomic configurations and electronic properties [4-7]. As a large-size reconstruction, theoretically predicting its atomic structure remains a significant challenge [10-17] in condensed matter physics for any traditional crystal structure prediction methods. Some candidate structures for the Si(111)-7×7 surface were initially proposed based on intuition and manually constructed [18-22]. The most famous one is the well-known dimer-adatom-stacking-fault model (DAS) proposed by K. Takayanagi and his co-workers in 1985 [23]. Only in recent years have papers been published on methods capable of directly predicting the DAS structure [15-17]. The STM image of DAS model has been simulated by many groups, which can help us to understand the experimental results well. Consequently, the DAS model has become widely recognized as the structural candidate for the Si(111)-7×7 reconstruction in further theoretical investigations [24-36].

From an electronic property perspective, the metallic nature of DAS is consistent with experimental results obtained by photoemission and scanning tunneling spectroscopy (**STS**) at room temperature [37-39], as well as some low-temperature photoemission measurements [40-42]. However, it contrasts with the insulating features [7, 43] detected in low-temperature photoemission spectra [37], as well as the presence of hard or soft energy gaps detected below 20 K in STS spectra [44-45] and inferred from charge transport experiments [46]. In order to continue using the metallic DAS mode, the soft energy gap is understood by long-range interaction induced V-shaped Coulomb gap [47] and the hard energy gap of about tens of meV is explained as the enhanced soft Coulomb gap due to local Coulomb blockade effects [44-46]. However, modified calculations such as DFT+U approximations and spin-split calculations also show that DAS is still a metallic phase [7]. Vertical displacements of adatoms had also been considered in previous calculations [7], but the calculated band structures can't explain the metal-insulator transition.

In principle, theoretical calculations are considered incorrect if they fail to agree with experimental results. For first-principles calculations, the PBE functional usually underestimates the split-energy of crystalline-field and underestimates the band gaps of semiconductors. This may one of the reasons why the DAS model fails to explain the insulating features reported in Si(111)-7×7 surface reconstruction. However, High-level methods such as HSE06 functionals [48] and GW approximations [49] are computationally too expensive to resolve this issue. Another possible reason is that the DAS model is not the only one structural candidate for reconstructed Si(111)-7×7 surface [50-52]. There may exist low-energy insulating phases that reconstruct either directly from the clean surface or as modifications of the DAS model. Recently, J. E. Demuth proposed a digitated faulted adatom (DFA) model as an alternative to the DAS model for explaining Si(111)-7×7 surface reconstruction [51-52]. But such a new DFA model has been quickly commented by other researchers due to its instability [53]. Thus, new reconstructed models are needed to address the unresolved issues in the Si(111)-7×7 surface. However, determining the atomic structure of such a large-scale reconstruction remains a significant challenge for traditional crystal structure prediction methods. Specialized and advanced techniques [15-17, 54] are required to tackle this problem.

In this work, the graph theory as implemented in $RG^2$ code [54] is employed to search possible reconstructions of Si(111)-7×7 surface. The $RG^2$ code has important advantages in handling large structure prediction [55] and has achieved numerous results in structural prediction [56-58]. Beyond the well-known DAS model of DAS-$d_8$-$T^{12}$, some low-energy reconstructions for Si(111)-7×7 surface has been successfully identified by graph theory in $RG^2$ code, including the DAS like models (DAS-$d_8$-$T^9H^{3\text{-}A}$, DAS-$d_8$-$T^9H^{3\text{-}B}$ and DAS-$d_8$-$T^6H^6$) and non-DAS models (AB-$d_{10}$-$T^{12}$, AB-$d_{10}$-$T^9H^3$, AA-$d_{10}$-$T^{12}$ and AA-$d_{10}$-$T^9H^3$), with and without stacking-fault, respectively. All these new models possess similar surface morphology to the famous DAS model (DAS-$d_8$-$T^{12}$), but with different atomic configurations in details. Their calculated surface energies, obtained from PBE-based first-principles calculations, are closely comparable, indicating similar stabilities. Furthermore, these models show STM images similar to that of the well-known DAS model (DAS-$d_8$-$T^{12}$). Although these reconstructions are metallic in NM state, they are further confirmed to be metals (DAS-d8-T9H3-B), half-metals (DAS-d8-T12, AB-d10-T9H3, AA-d10-T12 and AA-d10-T9H3), half-semimetals (DAS-d8-T9H3-A and DAS-d8-T6H6), or even insulators (AB-d10-T12), according to their FM band structures. Our work provides a plausible physical explanation for the metal–insulator transition observed on the Si(111) surface, and demonstrates the rich electromagnetic phases of reconstructed Si(111) surfaces for spintronic applications.

It is necessary to discuss the principles of reconstruction before searching new possible candidates. Symmetry breaking and changes in atomic coordination increase the surface energy, leading to reconstruction as the system seeks a lower-energy configuration. Two widely accepted factors play a crucial role in determining surface energy: chemical dangling bonds and geometrical distortions. A lower number of dangling bonds and minimal distortions generally correspond to lower surface energy. The clean Si(111) surface contains one dangling bond per unit cell (uc) and cannot undergo reconstruction due to the constraints imposed by periodicity and symmetry. In appropriately chosen supercells, such as the hexagonal √3×√3 cell, the addition of a single adatom at the tetragonal (T4) site can saturate all three dangling bonds on the surface. Although this introduces one additional dangling bond and creates three locally distorted atomic environments, the fully-covered √3×√3 supercell achieves a lower energy configuration. For larger supercells, such as 5×5 and 7×7, which cannot be fully covered by adatoms, additional dimerization of inner atoms must be considered. This mechanism, in cooperation with adatoms, leads to the formation of the well-known DAS (dimer-adatom-stacking fault) model, which further reduces the surface energy. However, DAS models can only be extended to (2n+1)×(2n+1) supercells, and the presence of interlayer stacking faults in discontinuous small triangular regions makes their dynamic and statistical mechanical properties difficult to understand [59].

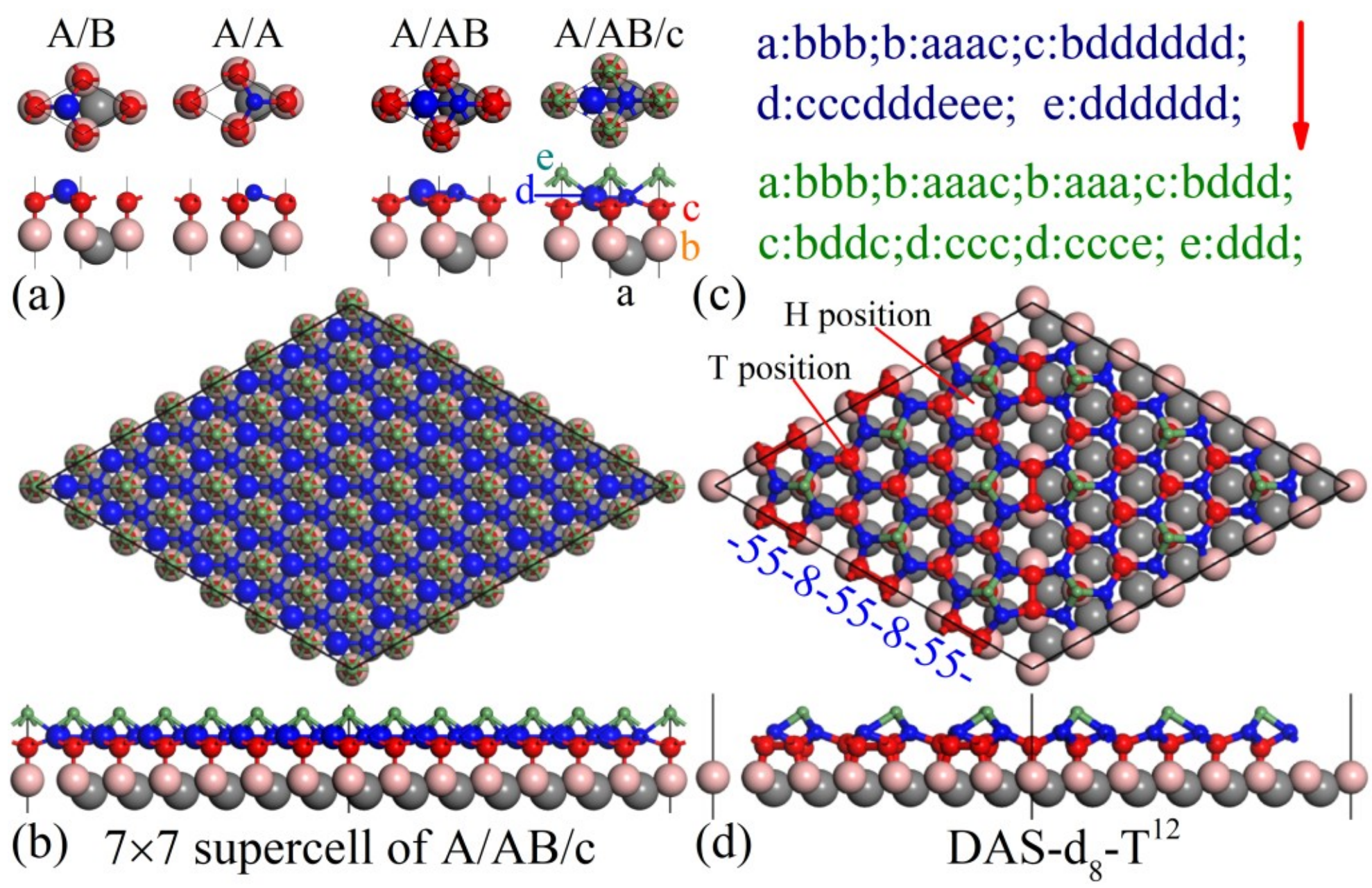

Fig.1. (a) The two fundamental stacking manners (A/A and A/B) in silicon bilayer and the hypothetical stacking manner without (A/AB) and with (A/AB/c) fully covered adatoms at the T4 (T) and hexagonal (H) positions. (b) the 7×7 supercell of A/AB/c model. (c) The adjacent relationships in the hypothetical A/AB/c (blue) and the reconstructed candidates (green). (d) The well-known DAS model for Si(111)-7×7 surface (DAS-$d_8$-$T^{12}$).

To find low energy reconstruction for Si(111)-7×7 surface, a graph-based strategy is proposed and implemented by $RG^2$ code[54]. As shown in Fig. 1 (a) are the two fundamental stacking orders in silicon-bilayers, A/A and A/B. Here, we hypothetically construct a hybridized one (A/AB) with fully covered adatoms in the T4 (T) and hexagonal (H) positions (A/AB/c), whose 7×7 supercell will be considered as starting point for reconstruction in $RG^2$. The five layers of atom in A/AB/c are labeled as a, b, c, d and e from inside to outside as shown in the picture, where the silicon layer formed by a and b are fixed for matching the bulk phase in the whole process. The adjacent relationships in this model can be easily featured as a:bbb, b:aaac, c:bdddddd, d:cccdddeee and e:dddddd, where a:bbb means that atom a connected to 3 atoms labeled as b. The principles to construct low-energy reconstructions can be summarized as follows: Remove as many surface atoms d as possible to reduce the number of dangling bonds. Create possible self-saturation between the inner atoms c that lose surface atoms. Retain as many adsorbed atoms e as possible to saturate the surface atoms d. Remove a small number of inner atoms c, translating the dangling bonds to the next inner layer. $RG^2$ can accomplish these tasks quickly. It will load the 7×7 supercell of A/AB/c and calculate its hexagonal symmetry subgroup (Nos.143 and 156) at the first step. And then, it randomly removes a small number of atoms labeled as c, as well as a large number of atoms labeled as d and e, under the constraints of subgroup symmetry. Finally, $RG^2$ will build the quotient graph (QG) for the rested framework and optimized the configurations with accepted adjacent relationships as listed in Fig1 (c). Those unlisted adjacent relationships are higher energy ones and not considered in this work. Here, a:bbb means atoms a with fixed positions for connecting bulk phase; b:aaac and b:aaa indicate the fixed atoms b for linking the surface layer, with and without new introduced dangling bonds, respectively; c:bddd and c:bddc are the adjacent relationships for atoms c in the surface layer with and without self-saturated dimerization, respectively; d:ccc and d:ccce indicate the rested atoms d in the outset surface layer with and without dangling bonds; And e:ddd means the rested adatoms e used for saturating the dangling bonds on the outset surface atoms d.

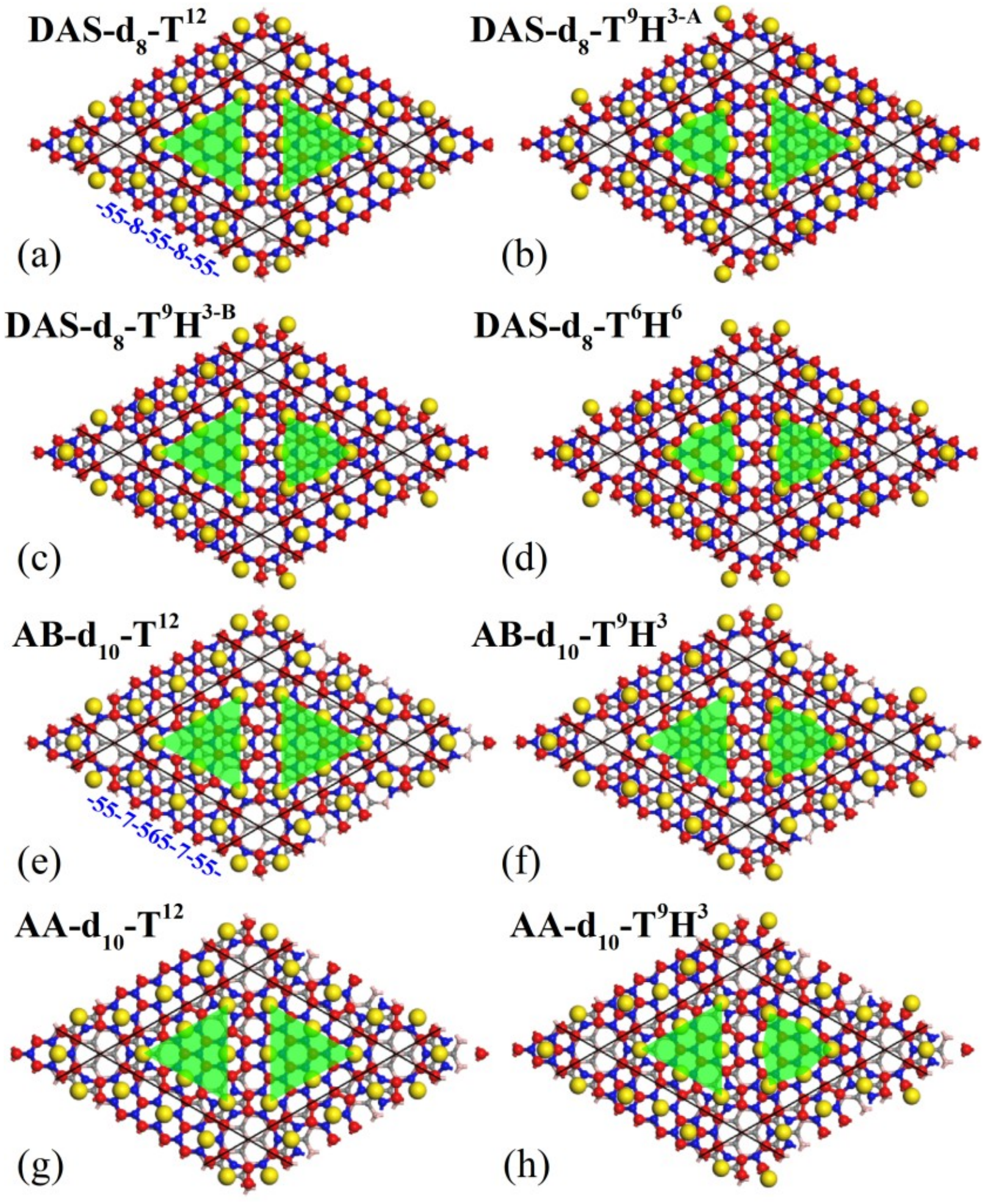

Fig.2. (a)-(d) are the DAS models DAS-$d_8$-$T^{12}$, DAS-$d_8$-$T^9H^{3\text{-}A}$, DAS-$d_8$-$T^9H^{3\text{-}B}$ and DAS-$d_8$-$T^6H^6$. (e)-(h) are the non-DAS models AB-$d_{10}$-$T^{12}$, AB-$d_{10}$-$T^9H^3$, AA-$d_{10}$-$T^{12}$ and AA-$d_{10}$-$T^9H^3$. Adatoms are marked as big yellow balls.

This approach is highly effective. It can rapidly reproduce the DAS-3×3 (within a few minutes) and DAS-5×5 (in about half an hour) reconstructions on a single machine. On a cluster with around 20 CPUs, the reconstruction of DAS-7×7 can be discovered after several hours of computation. Running this approach for an extended period on a cluster not only repeatedly finds DAS-3×3 and DAS-5×5 but also occasionally discovers DAS-7×7 and some previously unreported DAS and non-DAS reconstructions with and without stacking faults. The DAS and non-DAS models are featured by dimerization patters of -55-8-55-8-55- and -55-7-565-7-55- in Fig. 2(a) and (e), respectively. As summarized in Table S1 [60], all the DAS models (Fig. 2 (a)-(d), DAS-$d_8$-$T^{12}$, DAS-$d_8$-$T^9H^{3\text{-}A}$, DAS-$d_8$-$T^9H^{3\text{-}B}$ and DAS-$d_8$-$T^6H^6$) lost 8 atoms in reconstruction, including 1 inner half-layer atom c and 7 outer half-layer atoms d, respectively.

They contain 12 adatoms at the surface in different distributions of $T^{12}$, $T^9H^{3\text{-}A}$, $T^9H^{3\text{-}B}$ and $T^6H^6$, where T/H denotes the adatom positions and A/B indicates that H position is located at the AA/AB stacking area. Here, the well-known DAS model is cited as DAS-$d_8$-$T^{12}$ to distinguish it from other DAS structures, where $d_8$ and $T^{12}$ indicate that the lost 8 surface atoms and the 12 adatoms in the T4 positions. As the non-DAS models shown in Fig 2 (e)-(h), the reconstructed layer and the next inner layer exhibit a complete AB or AA stacking, rather than those with stacking-fault in mixed AA and AB manners. These non-DAS models also undergo the internal dimerization and surface passivation. They lost 10 surface atoms in reconstruction, including 1 inner half-layer atom c and 9 outer half-layer atoms d under the reconstruction. They possess also 12 adatoms in different distributions (AB-$d_{10}$-$T^{12}$, AB-$d_{10}$-$T^9H^3$, AA-$d_{10}$-$T^{12}$ and AA-$d_{10}$-$T^9H^3$), showing similar pictures to those in the DAS models. In table S1 [60], the dangling bonds (DB), pairs of dimerization (POD) and coverage rate ($R_c$) for different models of Si(111) surface are listed for comparison. The DAS models contain 9 pairs of self-saturated dimers (POD) and 19 total dangling bonds (tot-DB) distributing on the 12 adatoms (ad-DB), 6 unsaturated surface atoms (s-DB) and 1 inner-layer uncovered atom (iL-DB), respectively. However, the non-DAS models contain 12 PODs and 17 tot-DBs d istributing on the 12 adatoms, 4 unsaturated surface atoms, and 1 inner-layer uncovered atom.

Table 1. The summarize of the numbers for dangling bonds (ad-DB, s-DB, iL-DB and Tot-DB) and pairs of dimerization (POD) in different models, as well as the coverage rate ($R_c$) and surface energy ($E_f$) in unit of meV/Å$^2$ for different reconstructions in Si(111) surface.

| Systems | $E_{NM}$ (eV) | NM: $E_{surf}$ (meV/Å$^2$) | $E_{FM}$ (eV) | $E_{FM}$-$E_{FM}$ (meV) | FM: $E_{surf}$ (meV/Å$^2$) | Magmom (μB) | (FM) Electronic property |
|---|---|---|---|---|---|---|---|
| Clean 1×1 | -34.562 | 0 | -34.629 | -67 | 0 | 1.00 | Insulator |
| DAS-3×3 | -400.312 | -21.3 | -400.317 | -5 | -16.2 | 0.44 | - |
| DAS-5×5 | -1142873 | -23.8 | -1142.873 | 0 | -18.5 | 0 | - |
| DAS-$d_8$-$T^{12}$ | -2261.599 | -23.5 | -2261.609 | -10 | -18.3 | 1.00 | Half-Metal |
| DAS-$d_8$-$T^9H^{3\text{-}A}$ | -2259.787 | -20.7 | -2259.819 | -32 | -15.5 | 1.00 | Half-Semimetal |
| DAS-$d_8$-$T^9H^{3\text{-}B}$ | -2259.092 | -19.6 | -2259.137 | -45 | -14.4 | 1.22 | Metal |
| DAS-$d_8$-$T^6H^6$ | -2257.276 | -16.7 | -2257.311 | -35 | -11.5 | 1.00 | Half-Semimetal |
| AB-$d_{10}$-$T^{12}$ | -2246.457 | -16.8 | -2246.479 | -22 | -11.5 | 1.00 | Semiconductor |
| AB-$d_{10}$-$T^9H^3$ | -2243.902 | -12.8 | -2243.915 | -13 | -7.5 | 1.00 | Half-Metal |
| AA-$d_{10}$-$T^{12}$ | -2243.976 | -12.8 | -2243.996 | -20 | -7.6 | 1.00 | Half-Metal |
| AA-$d_{10}$-$T^9H^3$ | -2242.013 | -9.7 | -2242.021 | -8 | -4.5 | 1.00 | Half-Metal |

All these models are fully optimized under PBE functional as implemented in the widely used VASP code [61-63] based on a slab model including 3 bulk Si-layers covered by the reconstructed bilayer on the surface and hydrogen on the other side (Atom numbers are 445 and 443 for DAS and non-DAS models).

The lattice constants are fixed as those in the optimized bulk silicon as 27.069 Å. The convergence accuracy of energy and force are $10^{-5}$ eV and 0.01 eV/Å, respectively. The cutoff energy is set to be 400 eV and the k-mesh of 3×3×1 is used. In our work, we optimize the atomic positions under both ferromagnetic and non-magnetic states, and calculate the surface energy for each case separately. The surface energy ($E_{surf}$) can be evaluated as $E_{surf} = (E_{tot}^{rec-H} - E_{tot}^{1\times1-H} \times \frac{S_{tot}^{rec-H}}{S_{tot}^{1\times1-H}} - \mu \times N_{surface})/S_{tot}^{rec-H}$ [64-65], where $E_{tot}^{rec-H}$ *and* $S_{tot}^{rec-H}$ are the total energy and surface area of the reconstructed system, respectively, while $E_{tot}^{1\times1-H}$ *and* $S_{tot}^{1\times1-H}$ are the total energy and surface area of the clean surface in primitive cell (1×1). μ is the average energy calculated form bulk silicon and N accounts only the number of silicon atoms in the surface excluding the reference cell. As the results shown in Table 1, all reconstructions are exothermic and more stable than the clean surface (the 1×1-Ad0 considered as reference). These results indicate that all configurations favor the FM state, and the relative stability order among the structures remains unchanged between the FM and NM states. The surface energies under FM conditions are generally higher by approximately 4 meV/Å$^2$ compared to those under NM conditions. For the 7×7 reconstructions, the well-known DAS model of DAS-$d_8$-$T^{12}$ is confirmed to be the most stable one in both FM (-18.3 meV/Å$^2$) and NM (-23.5 meV/Å$^2$) states. And the following low-energies ones are DAS-$d_8$-$T^9H^{3-A}$, DAS-$d_8$-$T^6H^{6-B}$, AB-$d_{10}$-$T^{12}$, DAS-$d_8$-$T^9h^{3-B}$, AB-$d_{10}$-$T^9H^3$, AA-$d_{10}$-$T^{12}$ and AA-$d_{10}$-$T^9H^3$. It is noticed that the DAS-$d_8$-$T^6H^6$ and DAS-$d_8$-$T^9h^{3-B}$ have been reported in recent literature [17]. They were predicted possessing energies just 3.17 meV/Å$^2$ and 7.79 meV/Å$^2$ higher than DAS-$d_8$-$T^{12}$ in NM states, showing same energy order as our present work. The comparison and consistence between our results and those in previous literatures [9, 17] confirmed that our method and settings are reasonable for calculating Si(111) surface.

As shown in Fig.S1 [60] are the NM-state band structures of these models calculated from normal PBE functional. The results indicate that all these reconstructions are metallic phases with distinct energy bands crossing the Fermi level. It is worth noting that the number of bands crossing the Fermi level, their degree of isolation, and their occupancy differ among the structures. In the bands structures of DAS-$d_8$-$T^{12}$, DAS-$d_8$-$T^9H^{3-A}$, DAS-$d_8$-$T^6H^6$ and AA-$d_{10}$-$T^9H^3$, two partially overlapping yet energetically isolated bands cross the Fermi level (Fig.S2 (c)[60]), each being 3/4 filled. Systems DAS-$d_8$-$T^9H^{3-B}$, AB-$d_{10}$-$T^9H^3$ and AA-$d_{10}$-$T^{12}$ host three such isolated and overlapping bands filled to 5/6 (Fig.S2 (e) [60]), whereas system AB-$d_{10}$-$T^{12}$ exhibits a single half-filled isolated band intersecting the Fermi level (Fig S2 (b)). As indicated in Fig. S2 [60], these partially filled sets of isolated narrow bands tend to undergo spin splitting [66-69], which further lowers the total energy of the system. A phase transition from a NM metal to a FM insulator occurs at half-filling, while other filling levels lead to transitions into either FM metal, half-metal or half-semimetal.

To verify this trend, we calculated the band structures of these reconstructed FM states, as shown in Fig. 3. For the standard DAS model DAS-$d_8$-$T^{12}$, the isolated narrow band set (two isolated and overlapping bands) at 3/4 filling split into two sets—one fully occupied and the other half-filled—resulting in a typical half-metallic state. The other three configurations DAS-$d_8$-$T^9H^{3\text{-}A}$, DAS-$d_8$-$T^6H^6$ and AA-$d_{10}$-$T^9H^3$ are confirmed to be FM half-semimetal, half-metal and half-semimetal, respectively. DAS-$d_8$-$T^9H^{3\text{-}B}$, AB-$d_{10}$-$T^9H^3$ and AA-$d_{10}$-$T^{12}$ host 5/6-filled narrow band set (three isolated and overlapping bands) become FM metal, half-metal and half-metal. Indeed, only AB-$d^{10}$-$T^{12}$ exhibits a single half-filled isolated band crossing the Fermi level, leading to a transition into a FM insulator. These results are in excellent agreement with the above analysis presented in Fig. S2 [60]. As shown in Fig. S3 [60], such splitting behavior is also manifested in the band structures of the clean (111) surface in NM and FM states.

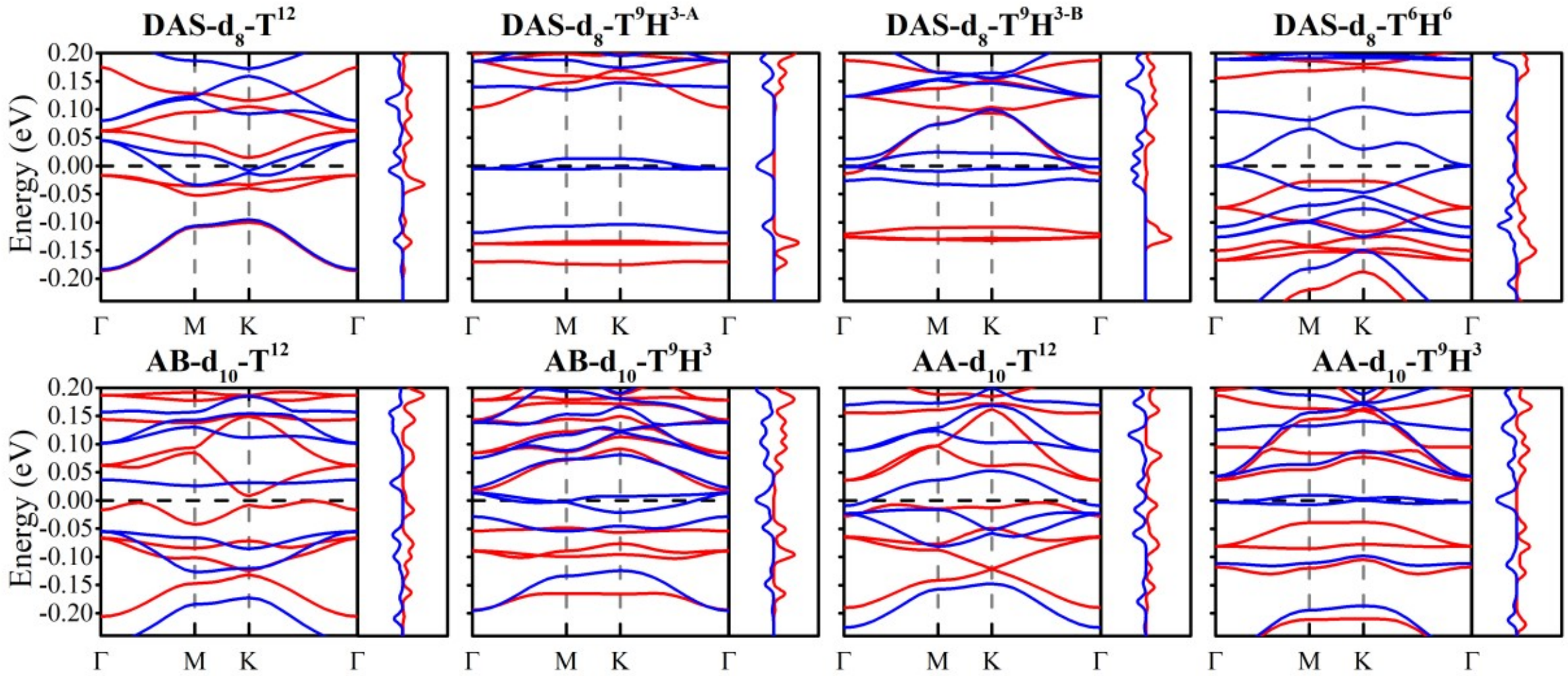

Fig.3 The band structures and density of states of various low-energy Si(111)-7×7 reconstructions calculated in FM states based on PBE functional.

Considering variations in the band structures—specifically the number and distribution of states below the Fermi level—we selected a broad energy window, including all states within 2 eV below the Fermi level. By adjusting the STM tip height, we obtained optimized images that clearly reveal all 12 bright spots simultaneously (Fig. 4). The resulting STM patterns are generally difficult to be distinguished without careful inspection. Among the DAS models, DAS-$d_8$-$T^{12}$ and DAS-$d_8$-$T^6H^6$ can be differentiated by their primary bright spots, forming two equilateral triangles and two Reuleaux triangles, respectively. In contrast, DAS-$d_8$-$T^9H^{3\text{-}A}$ and DAS-$d_8$-$T^9H^{3\text{-}B}$ each exhibit a mix of both shapes. For the non-DAS models (AB-$d_{10}$-$T^{12}$, AB-$d_{10}$-$T^9H^3$, AA-$d^{10}$-$T^{12}$, and AA-$d_{10}$-$T^9H^3$), the distribution of the 12 adatoms is easily distinguishable, though differentiating between AB and AA stacking in $T^{12}$ and $T^9H^3$ remains challenging. Notably, primary bright spots alone are insufficient to distinguish DAS-$d_8$-$T^{12}$ from AB-$d_{10}$-$T^{12}$ and AA-$d_{10}$-$T^{12}$, as all three display paired equilateral triangles. However, secondary bright spots provide a clue: AB-$d_{10}$-$T^{12}$ and AA-$d_{10}$-$T^{12}$ show a single-region triangular pattern, while DAS-$d_8$-$T^{12}$ exhibits six double-triangle features across two regions.

To provide more detailed STM images, we include supplementary STM patterns of all eight structures under various bias voltages and tip heights (see Fig.S4-S19 in the Supplementary Materials [60]), for future experimental comparison. Due to the large number of images, they are not discussed in detail in the main text. It should be noted that these images are not fully accurate, as they are based on PBE functional calculations that always underestimate the band gaps between bonding and anti-bonding states. In principle, bias shifts are expected under the more accurate HSE functional or GW calculations. However, due to the large size of these reconstructed surfaces, HSE and GW calculations are currently unfeasible. Therefore, the PBE-based results are provided here as a reference only.

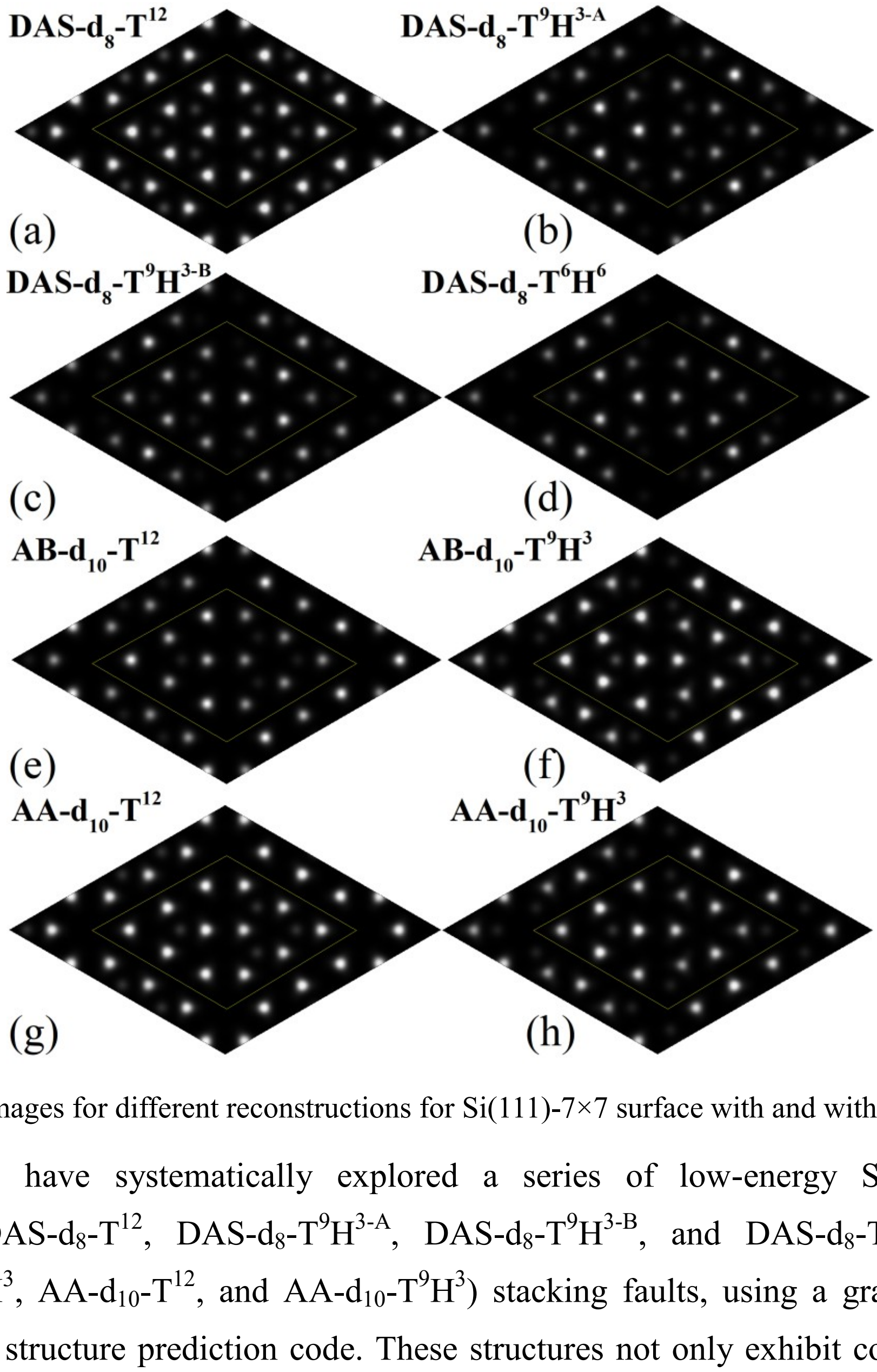

Fig.4 The simulated STM images for different reconstructions for Si(111)-7×7 surface with and without stacking fault.

In summary, we have systematically explored a series of low-energy Si(111)-7×7 surface reconstructions with (DAS-d$_8$-T$^{12}$, DAS-d$_8$-T$^9$H$^{3\text{-}A}$, DAS-d$_8$-T$^9$H$^{3\text{-}B}$, and DAS-d$_8$-T$^6$H$^6$) and without (AB-d$_{10}$-T$^{12}$, AB-d$_{10}$-T$^9$H$^3$, AA-d$_{10}$-T$^{12}$, and AA-d$_{10}$-T$^9$H$^3$) stacking faults, using a graph-based approach implemented in the RG2 structure prediction code. These structures not only exhibit comparable energetic stability and STM patterns to the well-established DAS model, but also capture the key features of the experimentally observed 7×7 reconstruction. All reconstructions are metallic in NM state, characterized by isolated narrow bands crossing the Fermi level with varying occupancy. Upon spin polarization, they transform into diverse electronic phases, including FM metals, half-metals, half-semimetals, and insulators, depending on their NM band occupations. These results highlight the rich electromagnetic phase behavior of Si(111) reconstructions and their relevance to spintronic applications. More importantly, the FM insulating phase of AB-$d_{10}$-$T^{12}$ offers a compelling physical picture for understanding the long-standing puzzle of the metal–insulator transition observed on the Si(111) surface.

## Acknowledgments

Chaoyu He acknowledges Prof. Chris J. Pickard for discussion about the calculation of surface energy, as well as Prof. J. E. Demuth for discussions on the DAS models and the issues within it. This work is supported by the National Natural Science Foundation of China (Grant Nos. 52372260, and 12204397, 12374046), the Youth Science and Technology Talent Project of Hunan Province (Grant No. 2022RC1197) and the Research Foundation of Education Bureau of Hunan Province, China (Grant No. 24A0121), the Science Fund for Distinguished Young Scholars of Hunan Province of China (No.2024JJ2048).

## Reference

[1] R. S. Becker, J. A. Golovchenko, and B. S. Swartzentruber, Tunneling Images of Germanium Surface Reconstructions and Phase Boundaries, Physical Review Letters 54, 2678 (1985).

[2] V. Cherepanov and B. Voigtländer, Influence of material, surface reconstruction, and strain on diffusion at the Ge(111) surface, Physical Review B 69, 125331 (2004).

[3] J. Schardt, J. Bernhardt, U. Starke, and K. Heinz, Crystallography of the (3×3) surface reconstruction of 3C-SiC(111), 4H-SiC(0001), and 6H-SiC(0001) surfaces retrieved by low-energy electron diffraction, Physical Review B 62, 10335 (2000).

[4] R. S. Becker, J. A. Golovchenko, D. R. Hamann, and B. S. Swartzentruber, Real-Space Observation of Surface States on Si(111) 7×7 with the Tunneling Microscope, Physical Review Letters 55, 2032 (1985).

[5] R. J. Hamers, R. M. Tromp, and J. E. Demuth, Surface Electronic Structure of Si(111)-(7×7) Resolved in Real Space, Physical Review Letters 56, 1972 (1986).

[6] R. E. Schlier and H. E. Farnsworth, Structure and Adsorption Characteristics of Clean Surfaces of Germanium and Silicon, The Journal of Chemical Physics 30, 917 (1959).

[7] S. Modesti, P. M. Sheverdyaeva, P. Moras, C. Carbone, M. Caputo, M. Marsi, E. Tosatti, and G. Profeta, Low-temperature insulating phase of the Si(111)-7×7 surface, Physical Review B 102, 035429 (2020).

[8] T. Eguchi and Y. Hasegawa, High Resolution Atomic Force Microscopic Imaging of the Si(111)-(7×7) Surface: Contribution of Short-Range Force to the Images, Physical Review Letters 89, 266105 (2002).

[9] R. A. Zhachuk and J. Coutinho, Crucial role of vibrational entropy in the Si(111)-7×7 surface structure stability, Physical Review B 105, 245306 (2022).

[10] I. Štich, M. C. Payne, R. D. King-Smith, J. S. Lin, and L. J. Clarke, Ab initio total-energy calculations for extremely large systems: Application to the Takayanagi reconstruction of Si(111), Physical Review Letters 68, 1351 (1992).

[11] K. D. Brommer, M. Needels, B. Larson, and J. D. Joannopoulos, Ab initio theory of the Si(111)-(7×7) surface reconstruction: A challenge for massively parallel computation, Physical Review Letters 68, 1355 (1992).

[12] S.-W. Wang, C.-R. Hsing, and C.-M. Wei, Expedite random structure searching using objects from Wyckoff positions, The Journal of Chemical Physics 148 (2018).

[13] M. K. Bisbo and B. Hammer, Efficient Global Structure Optimization with a Machine-Learned Surrogate Model, Physical Review Letters 124, 086102 (2020).

[14] M. N. Bauer, M. I. J. Probert, and C. Panosetti, Systematic Comparison of Genetic Algorithm and Basin Hopping Approaches to the Global Optimization of Si(111) Surface Reconstructions, The Journal of Physical Chemistry A 126, 3043 (2022).

[15] X. Du, J. K. Damewood, J. R. Lunger, R. Millan, B. Yildiz, L. Li, and R. Gómez-Bombarelli, Machine-learning-accelerated simulations to enable automatic surface reconstruction, Nature Computational Science 3, 1034 (2023).

[16] F. Brix, M.-P. Verner Christiansen, and B. Hammer, Cascading symmetry constraint during machine learning-enabled structural search for sulfur-induced Cu(111)-(43×43) surface reconstruction, The Journal of Chemical Physics 160 (2024).

[17] Y. Han, C. Ding, J. Wang, H. Gao, J. Shi, S. Yu, Q. Jia, S. Pan, and J. Sun, Efficient crystal structure prediction based on the symmetry principle, Nature Computational Science (2025).

[18] J. D. Levine, S. H. McFarlane, and P. Mark, Si (111) (7×7) surface structure: Calculations of LEED intensity and comparison with experiment, Physical Review B 16, 5415 (1977).

[19] M. J. Cardillo, Nature of the Si(111)7×7 reconstruction, Physical Review B 23, 4279 (1981).

[20] G. Binnig, H. Rohrer, C. Gerber, and E. Weibel, 7×7 Reconstruction on Si(111) Resolved in Real Space, Physical Review Letters 50, 120 (1983).

[21]P. A. Bennett, L. C. Feldman, Y. Kuk, E. G. McRae, and J. E. Rowe, Stacking-fault model for the Si(111)-(7×7) surface, Physical Review B 28, 3656 (1983).

[22] E. G. McRae, Surface stacking sequence and (7×7) reconstruction at Si(111) surfaces, Physical Review B 28, 2305 (1983).

[23] K. Takayanagi, Y. Tanishiro, M. Takahashi, and S. Takahashi, Structural analysis of Si(111)-7×7 by UHV-transmission electron diffraction and microscopy, Journal of Vacuum Science & Technology A 3, 1502 (1985).

[24] R. Zhachuk, B. Olshanetsky, J. Coutinho, and S. Pereira, Electronic effects in the formation of apparently noisy scanning tunneling microscopy images of Sr on Si(111)-7×7, Physical Review B 81, 165424 (2010).

[25] G.-X. Qian and D. J. Chadi, Si(111)-7×7 surface: Energy-minimization calculation for the dimer-adatom-stacking-fault model, Physical Review B 35, 1288 (1987).

[26] H. Lim, K. Cho, I. Park, J. D. Joannopoulos, and E. Kaxiras, Ab initio study of hydrogen adsorption on the Si(111)-(7×7) surface, Physical Review B 52, 17231 (1995).

[27] D. R. Alfonso, C. Noguez, D. A. Drabold, and S. E. Ulloa, First-principles studies of hydrogenated Si(111)-7×7, Physical Review B 54, 8028 (1996).

[28] J. Ortega, F. Flores, and A. L. Yeyati, Electron correlation effects in the Si(111)-7×7 surface, Physical Review B 58, 4584 (1998).

[29] F. J. Giessibl, S. Hembacher, H. Bielefeldt, and J. Mannhart, Subatomic Features on the Silicon (111)-(7×7) Surface Observed by Atomic Force Microscopy, 289, 422 (2000).

[30] R. Schillinger, C. Bromberger, H. J. Jänsch, H. Kleine, O. Kühlert, C. Weindel, and D. Fick, Metallic Si(111)-(7×7)-reconstruction: A surface close to a Mott-Hubbard metal-insulator transition, Physical Review B 72, 115314 (2005).

[31] J. A. McGuire, M. B. Raschke, and Y. R. Shen, Electron Dynamics of Silicon Surface States: Second-Harmonic Hole Burning on Si(111)-(7×7) Physical Review Letters 96, 087401 (2006).

[32] Y. Cho and R. Hirose, Atomic Dipole Moment Distribution of Si Atoms on a Si(111)-(7×7) Surface Studied Using Noncontact Scanning Nonlinear Dielectric Microscopy, Physical Review Letters 99, 186101 (2007).

[33] X.-Y. Ren, C.-Y. Niu, S. Yi, S. Li, and J.-H. Cho, Hydrogen adsorption induced nanomagnetism at the Si(111)-(7×7) surface, Physical Review B 98, 195424 (2018).

[34] A. P. Bartók, J. Kermode, N. Bernstein, and G. Csányi, Machine Learning a General-Purpose Interatomic Potential for Silicon, Physical Review X 8, 041048 (2018).

[35] Y. Shen, S. I. Morozov, K. Luo, Q. An, and W. A. Goddard Iii, Deciphering the Atomistic Mechanism of Si(111)-7×7 Surface Reconstruction Using a Machine-Learning Force Field, Journal of the American Chemical Society 145, 20511 (2023).

[36] L. Hu, B. Huang, and F. Liu, Atomistic Mechanism Underlying the Si(111)-(7×7) Surface Reconstruction Revealed by Artificial Neural-Network Potential, Physical Review Letters 126, 176101 (2021).

[37] R. I. G. Uhrberg, T. Kaurila, and Y. C. Chao, Low-temperature photoemission study of the surface electronic structure of Si(111)7×7, Physical Review B 58, R1730 (1998).

[38] P. Martensson, W.-X. Ni, G. V. Hansson, J. M. Nicholls, and B. Reihl, Surface electronic structure of Si(111)7×7-Ge and Si(111)5×5-Ge studied with photoemission and inverse photoemission, Physical Review B 36, 5974 (1987).

[39] J. Mysliveček, A. Stróżecka, J. Steffl, P. Sobotík, I. Ošt’ádal, and B. Voigtländer, Structure of the adatom electron band of the Si(111)-7×7 surface, Physical Review B 73, 161302 (2006).

[40] J. E. Demuth, B. N. J. Persson, and A. J. Schell-Sorokin, Temperature-Dependent Surface States and Transitions of Si(111)-7×7, Physical Review Letters 51, 2214 (1983).

[41] R. Losio, K. N. Altmann, and F. J. Himpsel, Fermi surface of Si(111)7×7, Physical Review B 61, 10845 (2000).

[42] I. Barke, F. Zheng, A. R. Konicek, R. C. Hatch, and F. J. Himpsel, Electron-Phonon Interaction at the Si(111)-7×7 Surface, Physical Review Letters 96, 216801 (2006).

[43] M. E. Dávila, J. Ávila, I. R. Colambo, D. B. Putungan, D. P. Woodruff, and M. C. Asensio, New insight on the role of localisation in the electronic structure of the Si(111)(7×7) surfaces, Scientific Reports 11, 15034 (2021).

[44] S. Modesti, H. Gutzmann, J. Wiebe, and R. Wiesendanger, Correction of systematic errors in scanning tunneling spectra on semiconductor surfaces: The energy gap of Si(111)-7×7 at 0.3 K, Physical Review B 80, 125326 (2009).

[45] A. B. Odobescu and S. V. Zaitsev-Zotov, Energy gap revealed by low-temperature scanning–tunnelling spectroscopy of the Si(111)-7×7 surface in illuminated slightly doped crystals, Journal of Physics: Condensed Matter 24, 395003 (2012).

[46] A. B. Odobescu, A. A. Maizlakh, and S. V. Zaitsev-Zotov, Electron correlation effects in transport and tunneling spectroscopy of the Si(111)-7×7 surface, Physical Review B 92, 165313 (2015).

[47] A. L. Efros and B. I. Shklovskii, Coulomb gap and low temperature conductivity of disordered systems, Journal of Physics C: Solid State Physics 8, L49 (1975).

[48] J. Heyd, G. E. Scuseria, and M. Ernzerhof, Hybrid functionals based on a screened Coulomb potential, The Journal of Chemical Physics 118, 8207 (2003).

[49] L. Hedin, New Method for Calculating the One-Particle Green's Function with Application to the Electron-Gas Problem, Physical Review 139, A796 (1965).

[50] J. E. Demuth, Experimental Evidence for a New Two-Dimensional Honeycomb Phase of Silicon: A Missing Link in the Chemistry and Physics of Silicon Surfaces?, The Journal of Physical Chemistry C 124, 22435 (2020).

[51] J. E. Demuth, Evidence for Symmetry Breaking in the Ground State of the Si(111)-7×7: The Origin of Its Metal–Insulator Transition, physica status solidi (b) 257, 2000229 (2020).

[52] J. E. Demuth, A re-evaluation of diffraction from Si(111) 7 × 7: decoding the encoded phase information in the 7 × 7 diffraction pattern, Physical Chemistry Chemical Physics 23, 8043 (2021).

[53] R. A. Zhachuk and J. Coutinho, Comment on "Experimental Evidence for a New Two-Dimensional Honeycomb Phase of Silicon: A Missing Link in the Chemistry and Physics of Silicon Surfaces?", The Journal of Physical Chemistry C 126, 866 (2022).

[54] X. Shi, C. He, C. J. Pickard, C. Tang, and J. Zhong, Stochastic generation of complex crystal structures combining group and graph theory with application to carbon, Physical Review B 97, 014104 (2018).

[55] C. He, X. Shi, S. J. Clark, J. Li, C. J. Pickard, T. Ouyang, C. Zhang, C. Tang, and J. Zhong, Complex Low Energy Tetrahedral Polymorphs of Group IV Elements from First Principles, Physical Review Letters 121, 175701 (2018).

[56] S. Li, X. Shi, J. Li, C. He, T. Ouyang, C. Tang, and J. Zhong, Spin-orbital coupling induced isolated flat band in bismuthene with k-dependent spin texture, Physical Review B 110, 115115 (2024).

[57] C. He, S. Li, Y. Zhang, Z. Fu, J. Li, and J. Zhong, Isolated zero-energy flat bands and intrinsic magnetism in carbon monolayers, Physical Review B 111, L081404 (2025).

[58] C. He, Y. Liao, T. Ouyang, H. Zhang, H. Xiang, and J. Zhong, Flat-band based high-temperature ferromagnetic semiconducting state in the graphitic C4N3 monolayer, Fundamental Research 5, 138 (2025).

[59] M. Smeu, H. Guo, W. Ji, and R. A. Wolkow, Electronic properties of Si(111)-7×7 and related reconstructions: Density functional theory calculations, Physical Review B 85, 195315 (2012).

[60] See Supplemental Material at [URL] for computational details, optimized crystal information files, as well as the supplemental figures and tables.

[61] G. Kresse and J. Furthmüller, Efficient iterative schemes for ab initio total-energy calculations using a plane-wave basis set, Physical Review B 54, 11169 (1996).

[62] J. P. Perdew, K. Burke, and M. Ernzerhof, Generalized Gradient Approximation Made Simple, Physical Review Letters 77, 3865 (1996).

[63] G. Kresse and D. Joubert, From ultrasoft pseudopotentials to the projector augmented-wave method, Physical Review B 59, 1758 (1999).

[64] R. Zhachuk, S. Teys, and J. Coutinho, Strain-induced structure transformations on Si(111) and Ge(111) surfaces: A combined density-functional and scanning tunneling microscopy study, The Journal of Chemical Physics 138 (2013).

[65] R. Zhachuk, J. Coutinho, A. Dolbak, V. Cherepanov, and B. Voigtländer, Si(111) strained layers on Ge(111): Evidence for c(2×4) domains, Physical Review B 96, 085401 (2017).

[66] Chaoyu He, Yujie Liao, Tao Ouyang, Huimin Zhang, Hongjun Xiang, Jianxin Zhong,Flat-band based high-temperature ferromagnetic semiconducting state in the graphitic C4N3 monolayer, Fundamental Research 5, 138 (2025) .

[67]Chaoyu He, Shifang Li, Yuwen Zhang, Zhentao Fu, Jin Li and Jianxin Zhong, Isolated zero-energy flat bands and intrinsic magnetism in carbon monolayers, Physical Review B 111, L081404 (2025)

[68] M. Maksymenko, A. Honecker, R. Moessner, J. Richter and O. Derzhko, Flat-band ferromagnetism as a Pauli-correlated percolation problem, Physical Review Letter 109, 096404 (2012).

[69] E. Bobrow, J. Zhang, Y. Li, Ferromagnetic percolation transition in a multiorbital flat band assisted by Hund's coupling, Physical Review B 104, 064442 (2021).

# Supplementary for "First-principles predictions of the diversity in atomic structures and electronic properties of the reconstructed Si(111)-7×7 surface"

Table S1. The summarize of the numbers for dangling bonds (ad-DB, s-DB, iL-DB and Tot-DB) and pairs of dimerization (POD) in different models, as well as the coverage rate ($R_c$) for different reconstructions in Si(111) surface.

| Systems | ad-DB | s-DB | iL-DB | tot-DB | POD | $R_c$ | $N_H$ | $N_{Si}$ | $N_{tot}$ |
|---|---|---|---|---|---|---|---|---|---|
| DAS-3×3 | 2 | 0 | 1 | **3** | 3 | 4/9 | 9 | 70 | 79 |
| DAS-5×5 | 6 | 2 | 1 | **9** | 6 | 6/25 | 25 | 200 | 225 |
| DAS-$d_8$-$T^{12}$ | 12 | 6 | 1 | **19** | 9 | 12/49 | 49 | 394 | 443 |
| DAS-$d_8$-$T^9H^{3\text{-}B}$ | 12 | 6 | 1 | **19** | 9 | 12/49 | 49 | 394 | 443 |
| DAS-$d_8$-$T^9H^{3\text{-}A}$ | 12 | 6 | 1 | **19** | 9 | 12/49 | 49 | 394 | 443 |
| DAS-$d_8$-$T^6H^6$ | 12 | 6 | 1 | **19** | 9 | 12/49 | 49 | 394 | 443 |
| AB-$d_{10}$-$T^{12}$ | 12 | 4 | 1 | **17** | 12 | 12/49 | 49 | 396 | 445 |
| AB-$d_{10}$-$T^9H^3$ | 12 | 4 | 1 | **17** | 12 | 12/49 | 49 | 396 | 445 |
| AA-$d_{10}$-$T^{12}$ | 12 | 4 | 1 | **17** | 12 | 12/49 | 49 | 396 | 445 |
| AA-$d_{10}$-$T^9H^3$ | 12 | 4 | 1 | **17** | 12 | 12/49 | 49 | 396 | 445 |

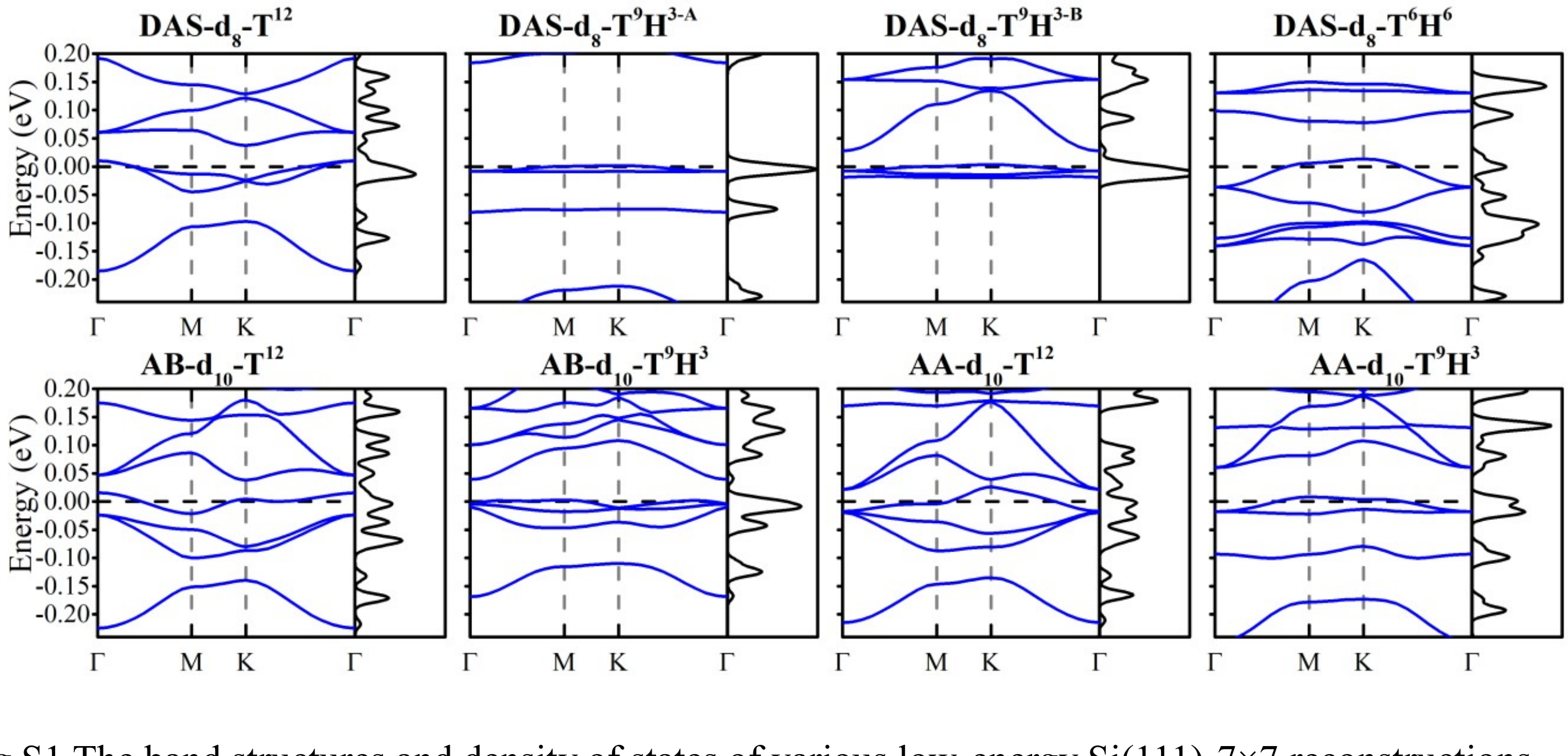

Fig.S1 The band structures and density of states of various low-energy Si(111)-7×7 reconstructions calculated in NM states based on PBE functional.

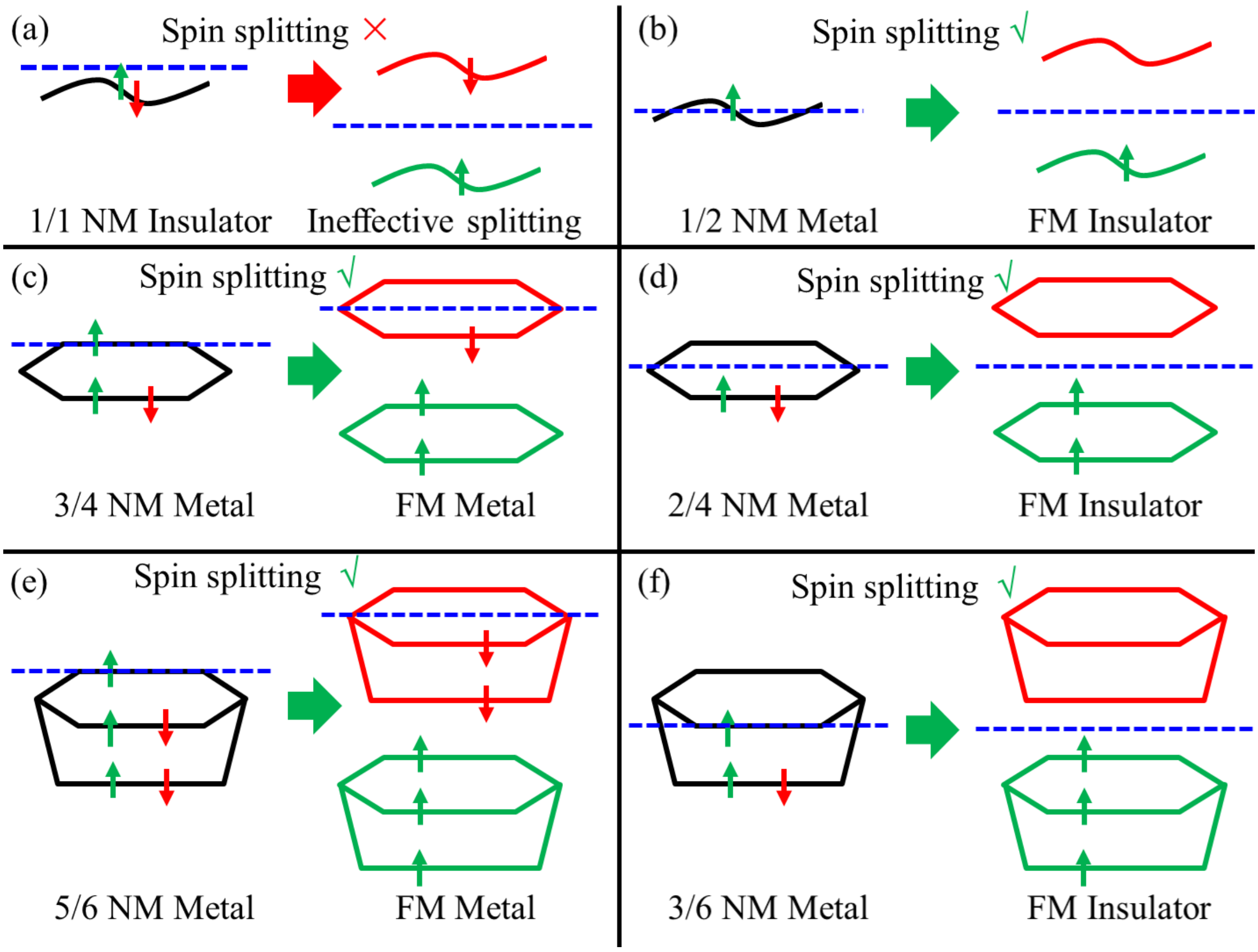

Fig.S2 Schematic illustration of spin splitting in isolated narrow-band sets under different filling conditions. Colored solid lines represent energy bands (black for non-spin-polarized, red and green for spin-up and spin-down, respectively). The blue dashed line indicates the Fermi-level, and the up/down arrows represent electron occupancy.

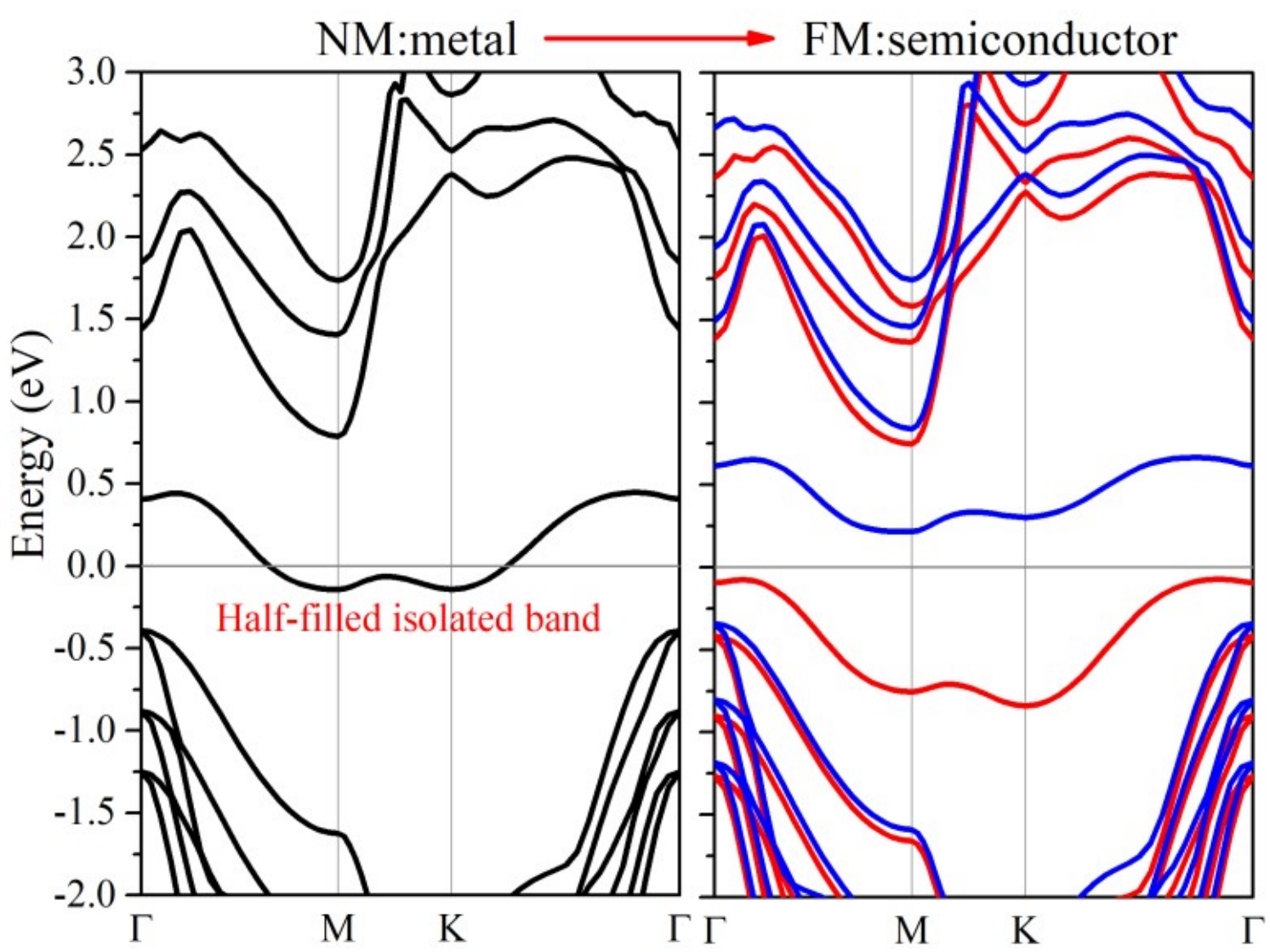

Fig.S3 Band structures of the clean Si(111) surface under different magnetic states

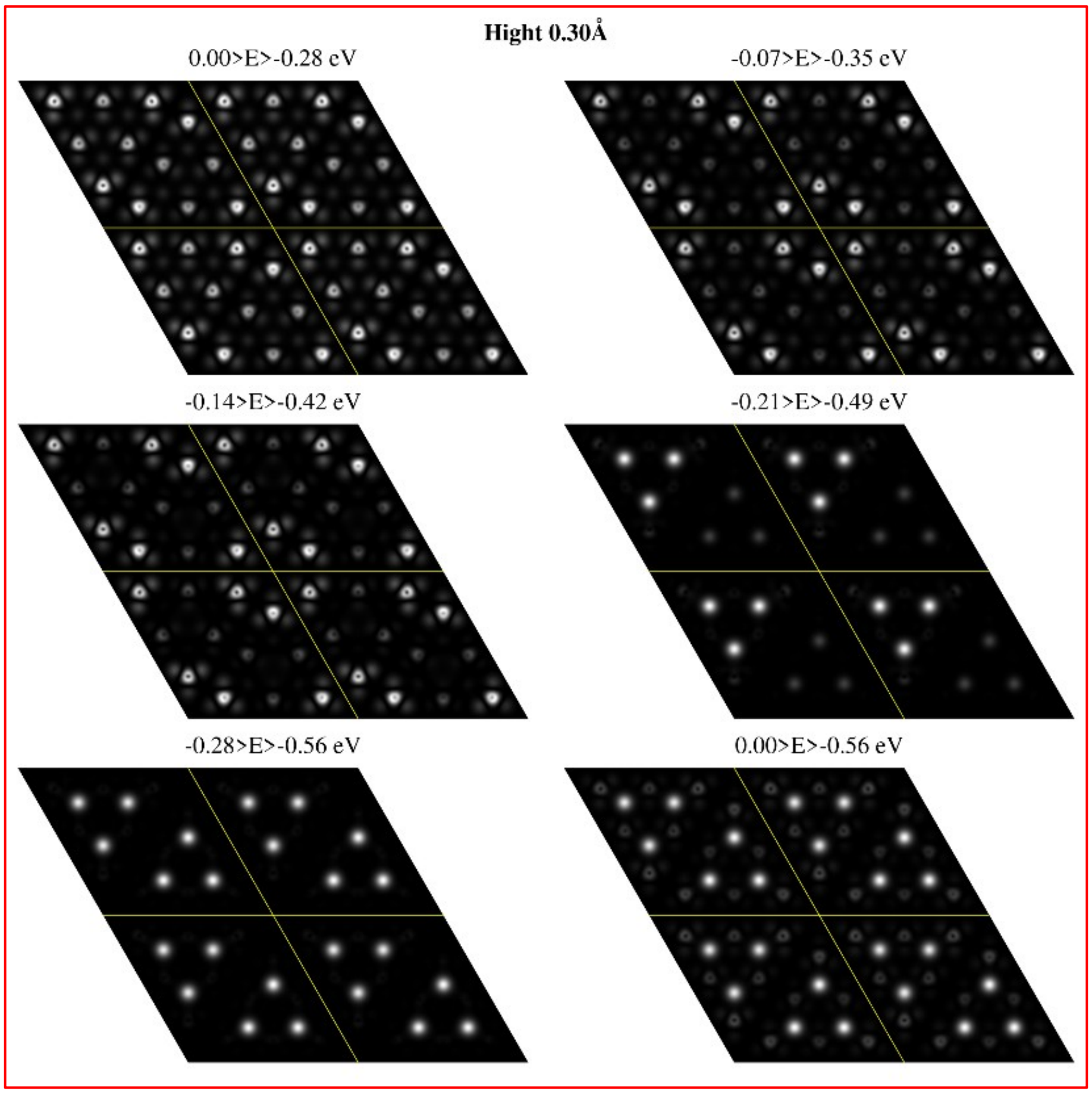

Fig.S4 STM of DAS-d8-T12 at the hight of 0.3 Å

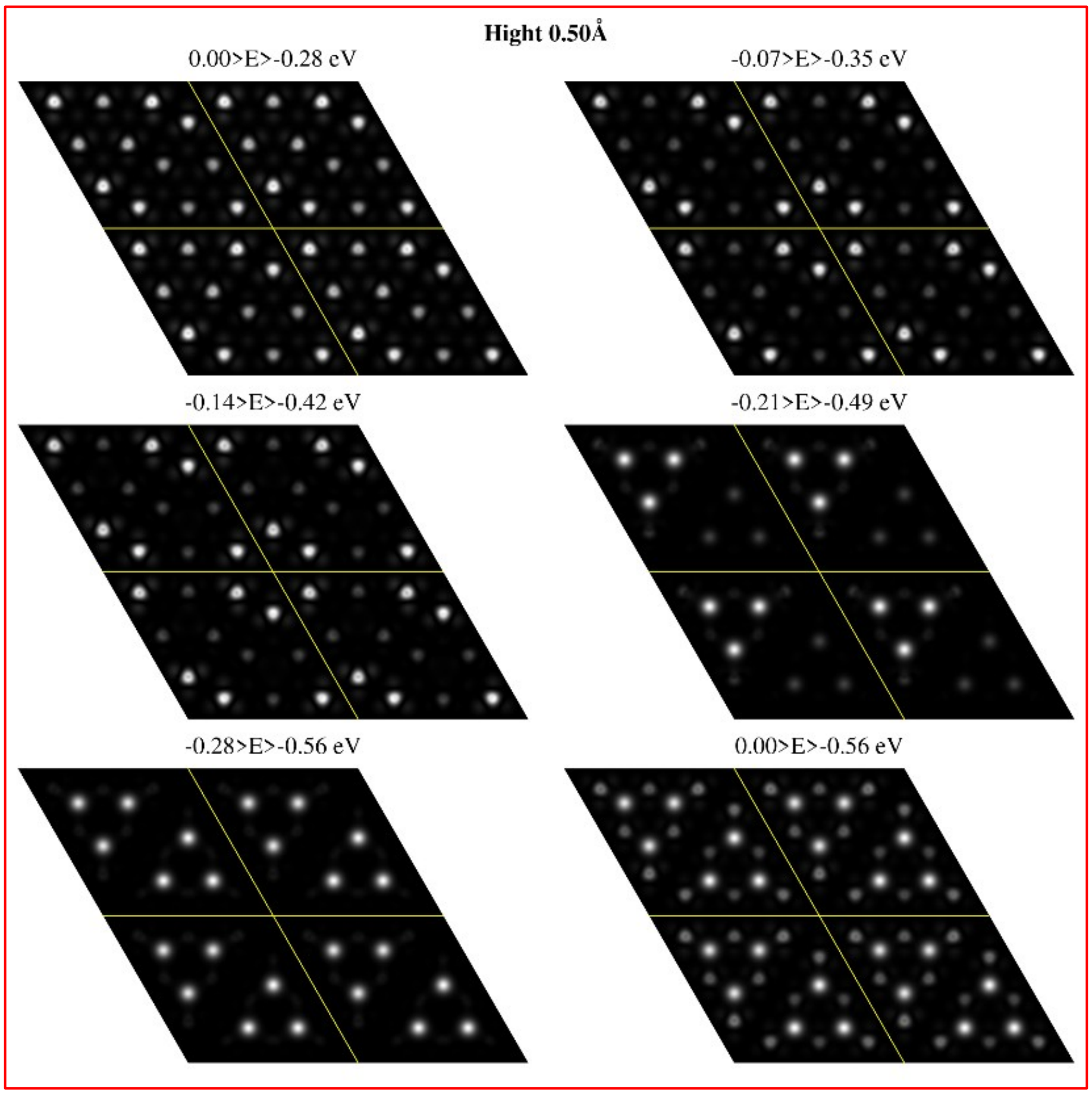

Fig.S5 STM of DAS-d8-T12 at the hight of 0.5 Å

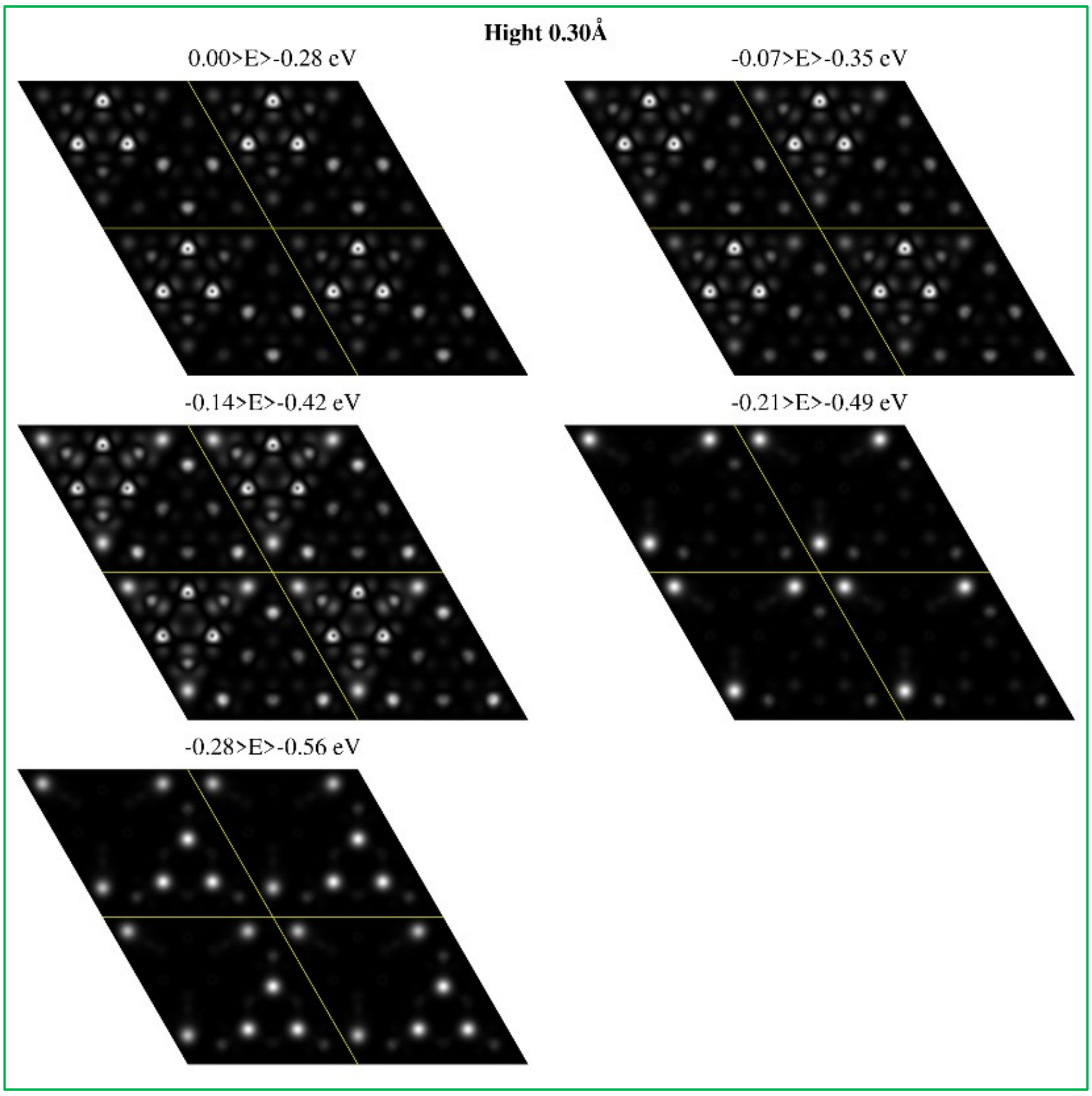

Fig.S6 STM of DAS-d8-T9H3-A at the hight of 0.3 Å

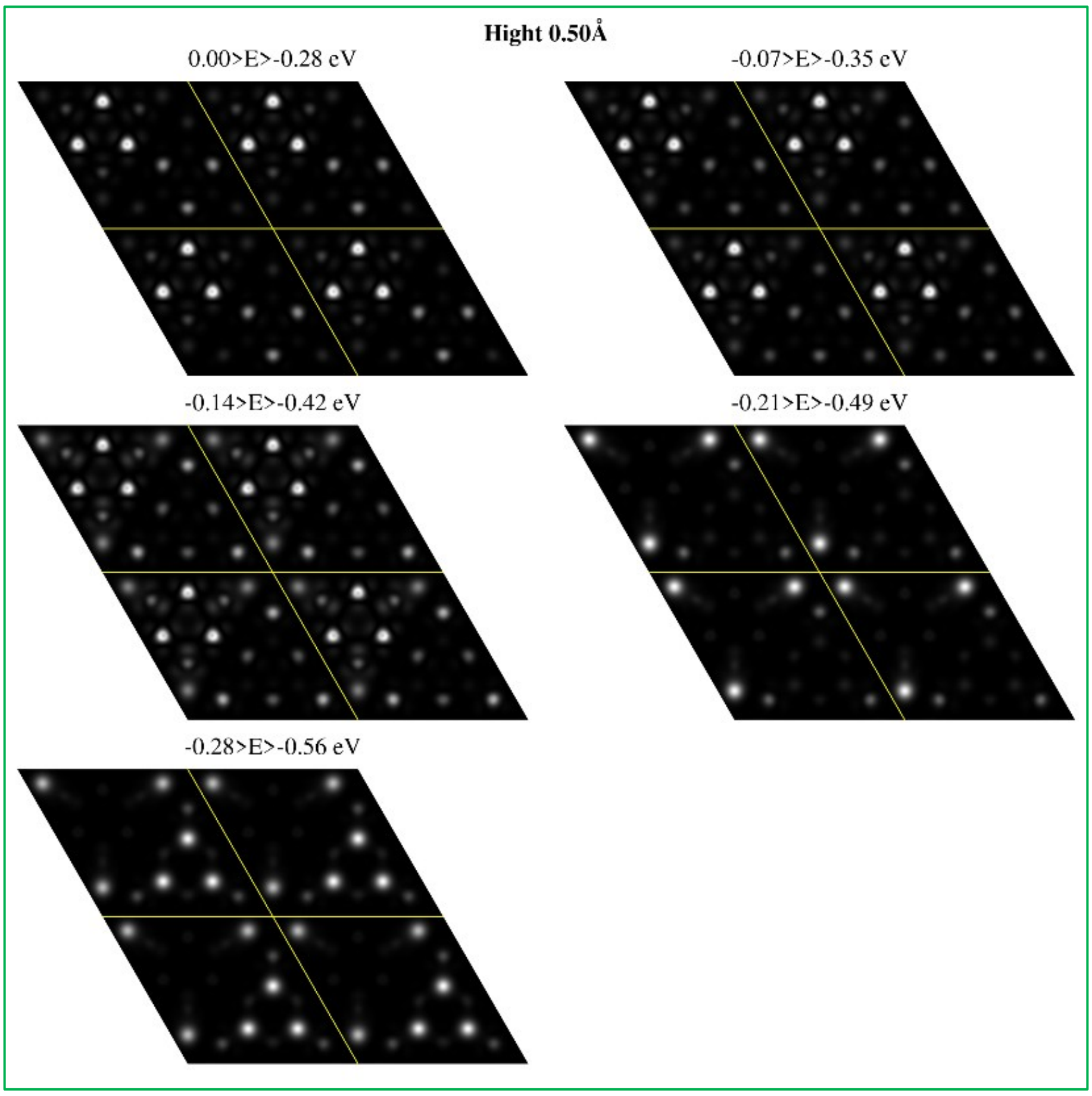

Fig.S7 STM of DAS-d8-T9H3-A at the hight of 0.5 Å

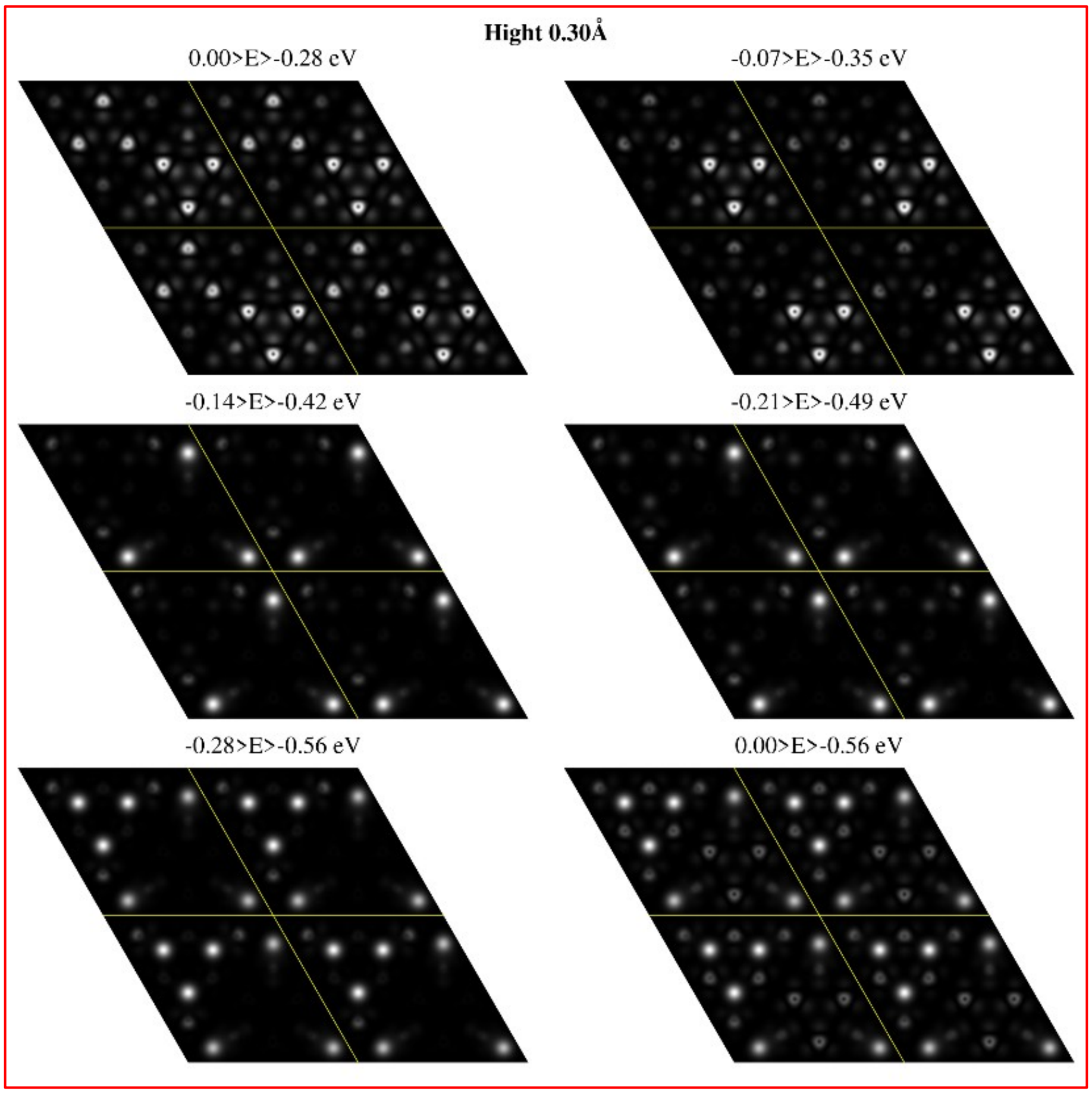

Fig.S8 STM of DAS-d8-T9H3-B at the hight of 0.3 Å

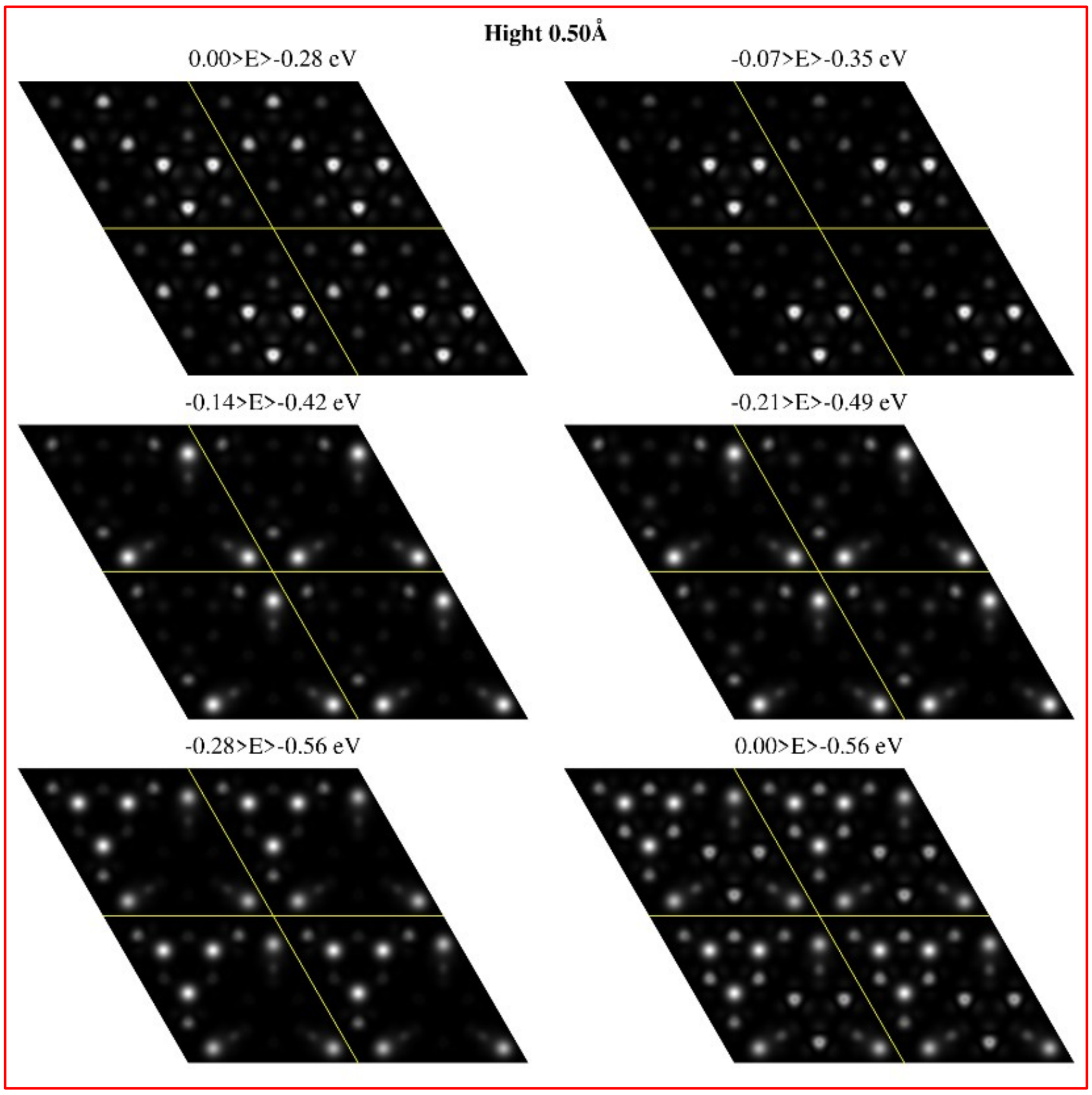

Fig.S9 STM of DAS-d8-T9H3-B at the hight of 0.5 Å

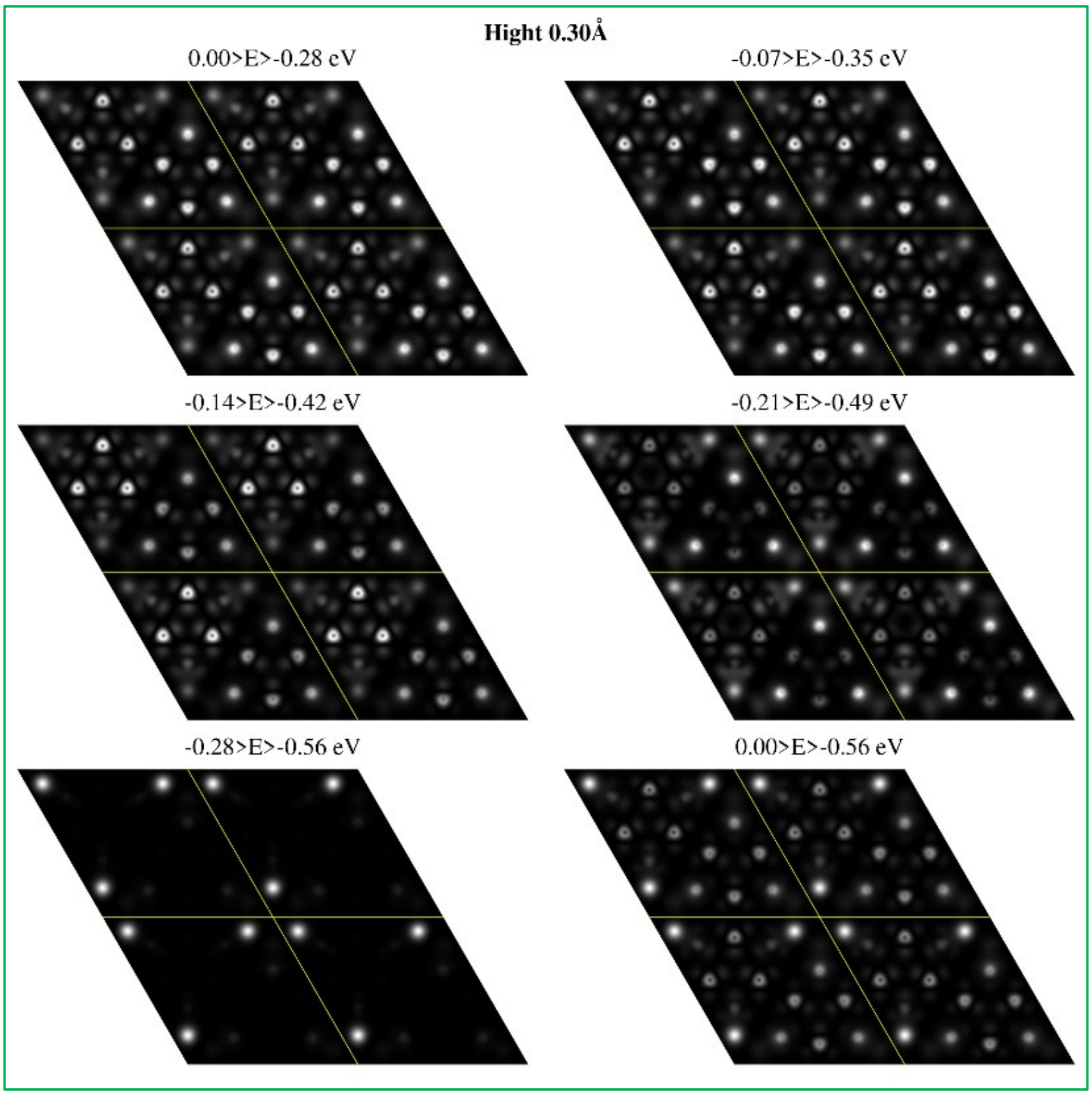

Fig.S10 STM of DAS-d8-T6H6 at the hight of 0.3 Å

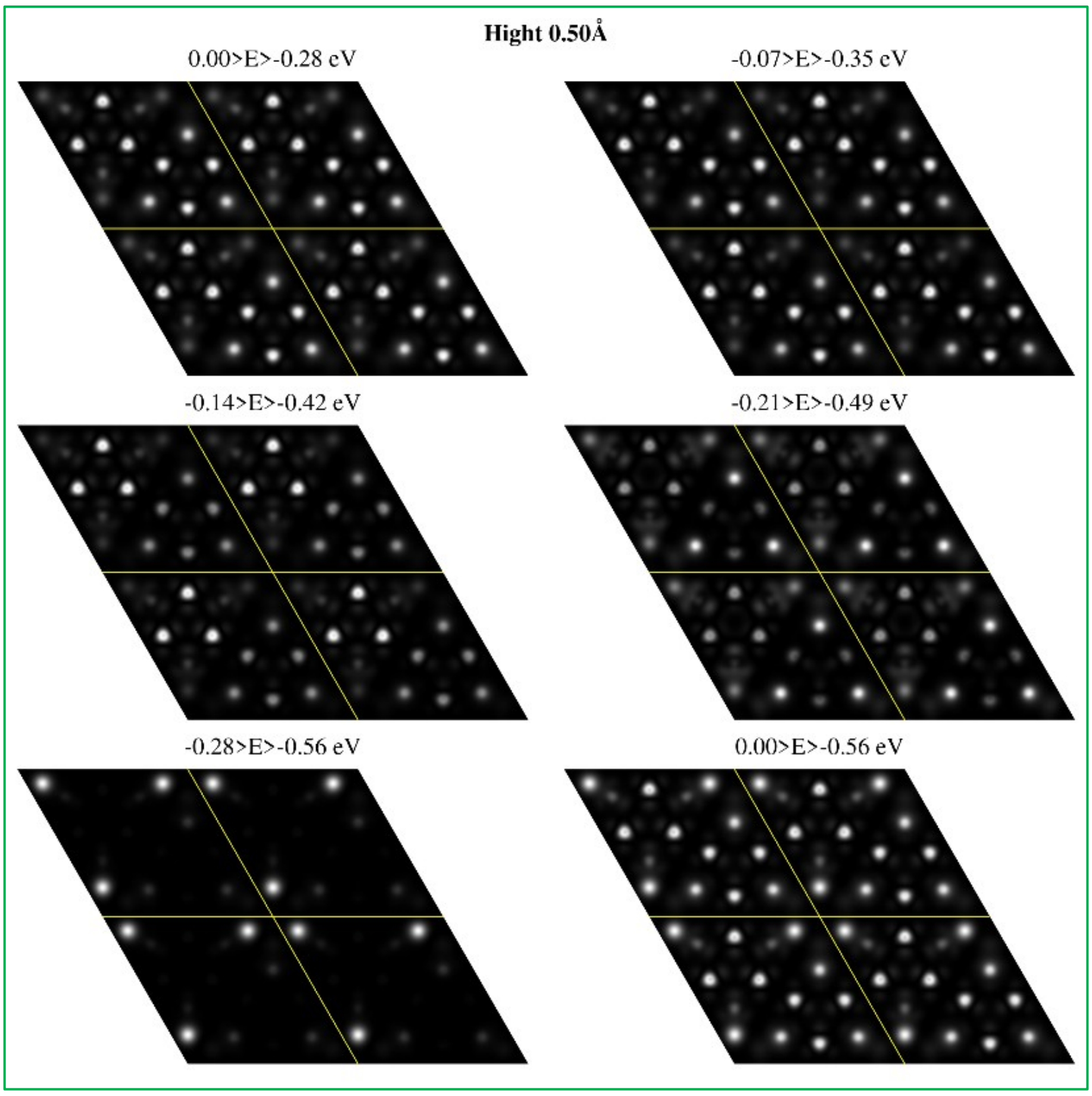

Fig.S11 STM of DAS-d8-T6H6 at the hight of 0.5 Å

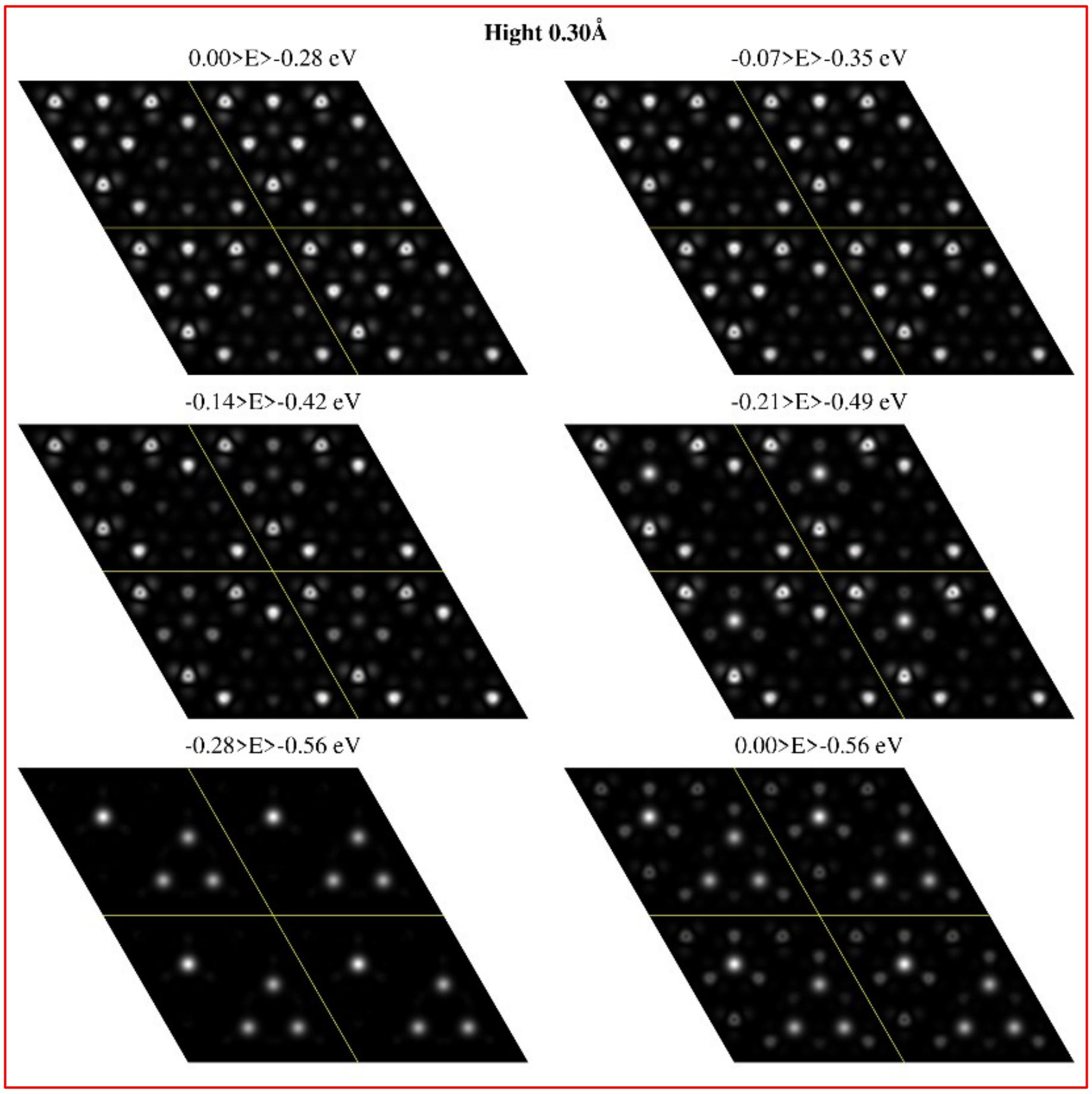

Fig.S12 STM of AB-d10-T12 at the hight of 0.3 Å

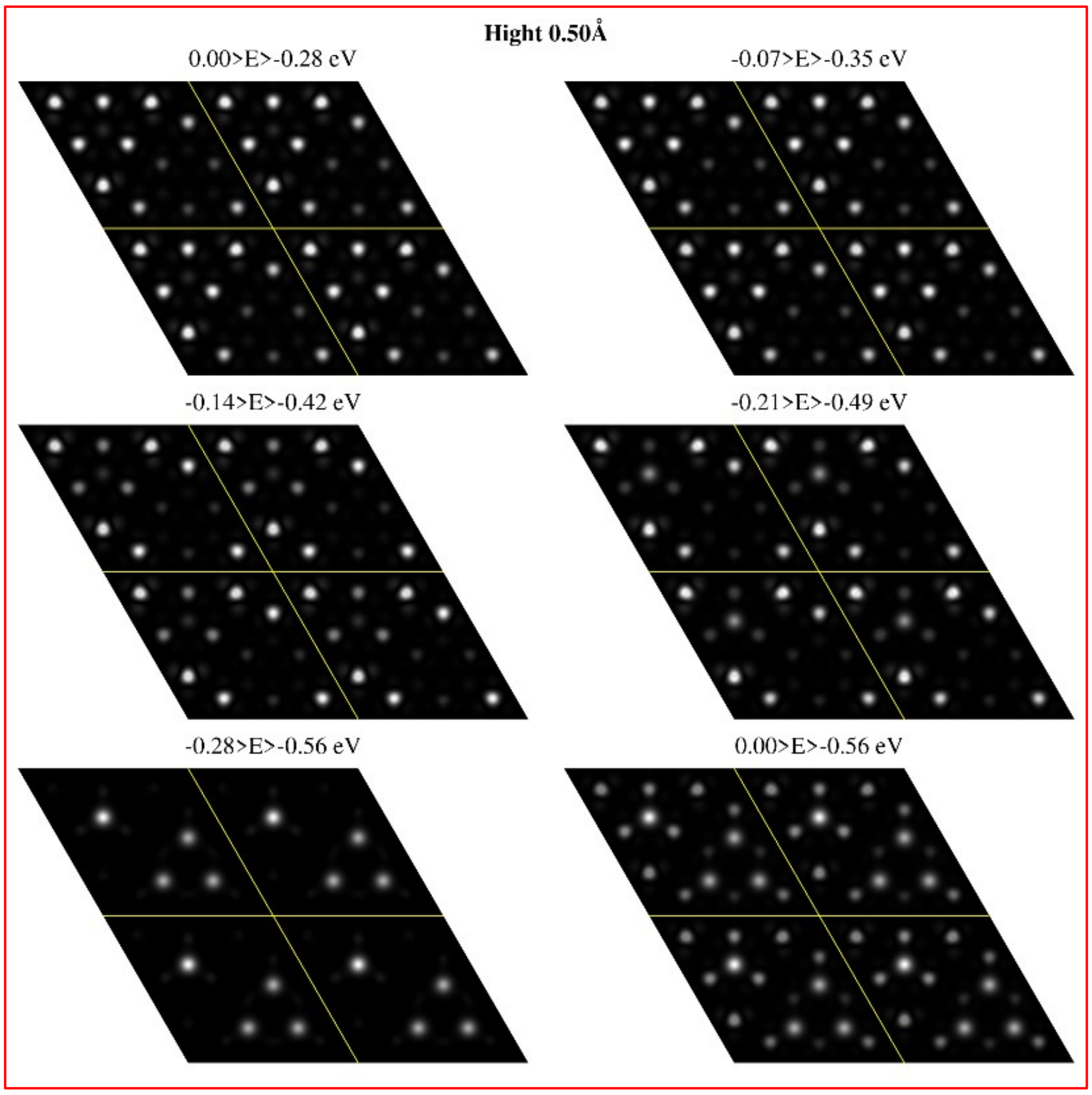

Fig.S13 STM of AB-d10-T12 at the hight of 0.5Å

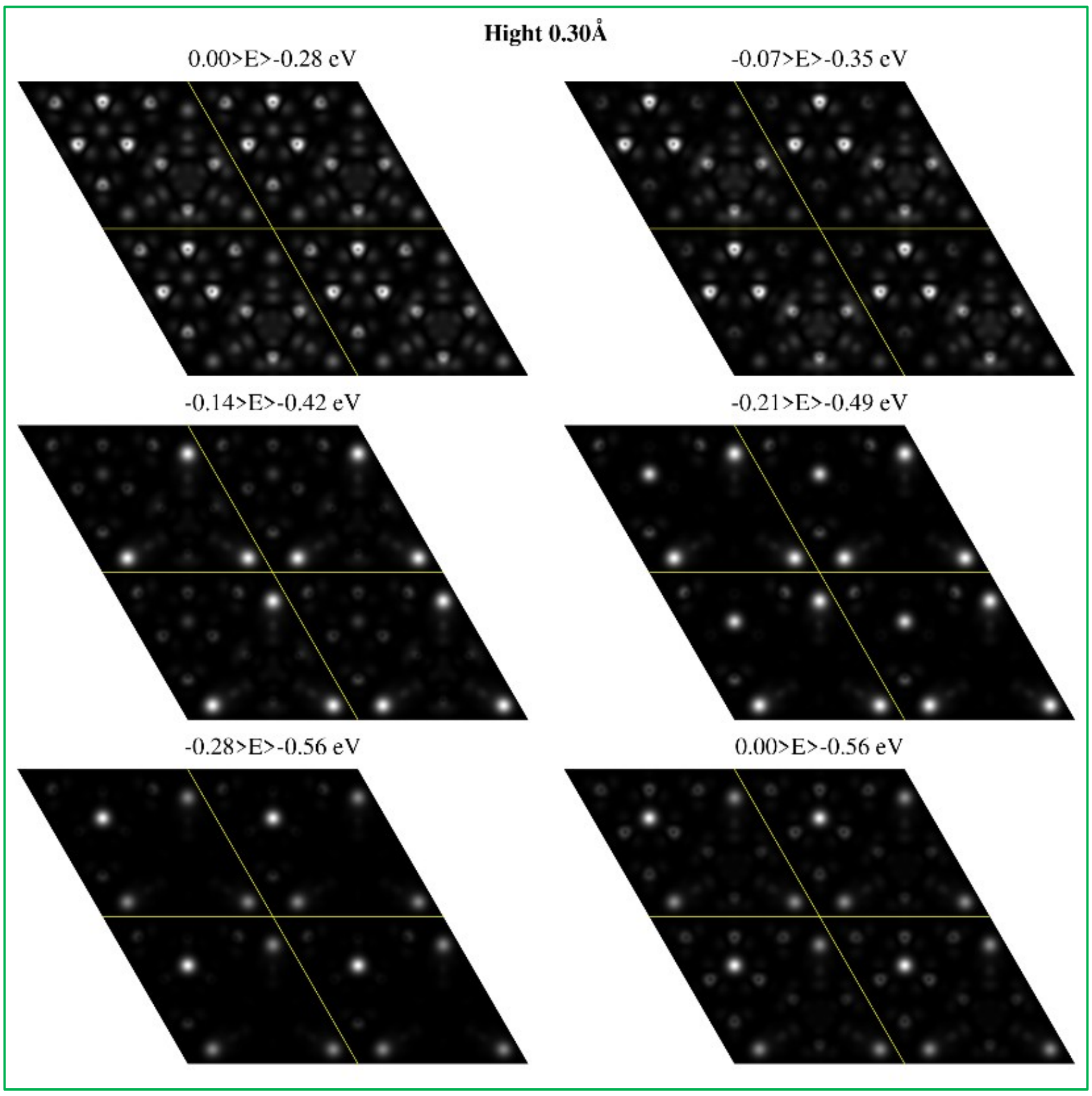

Fig.S14 STM of AB-d10-T9H3 at the hight of 0.3 Å

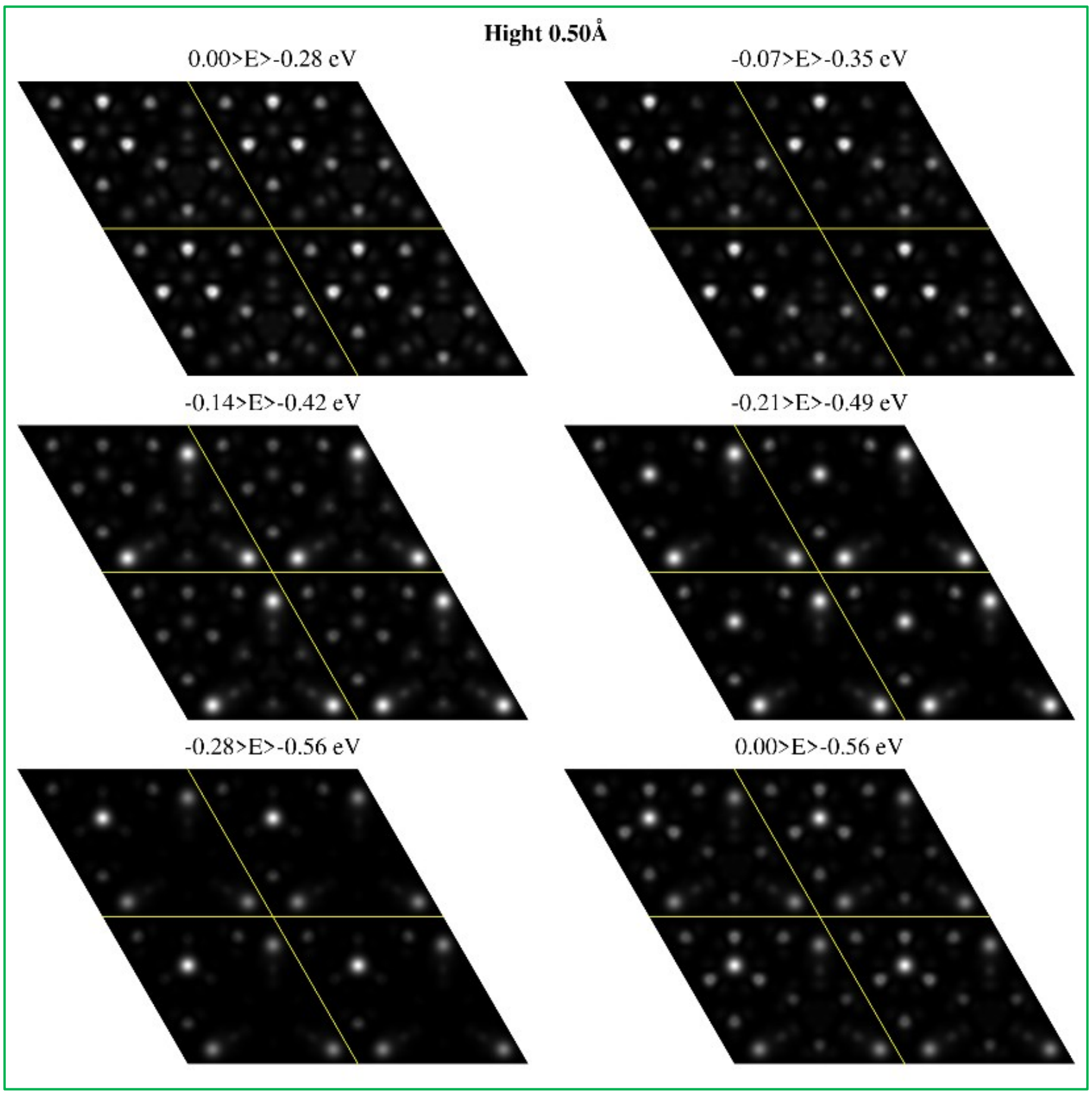

Fig.S15 STM of AB-d10-T9H3 at the hight of 0.5 Å

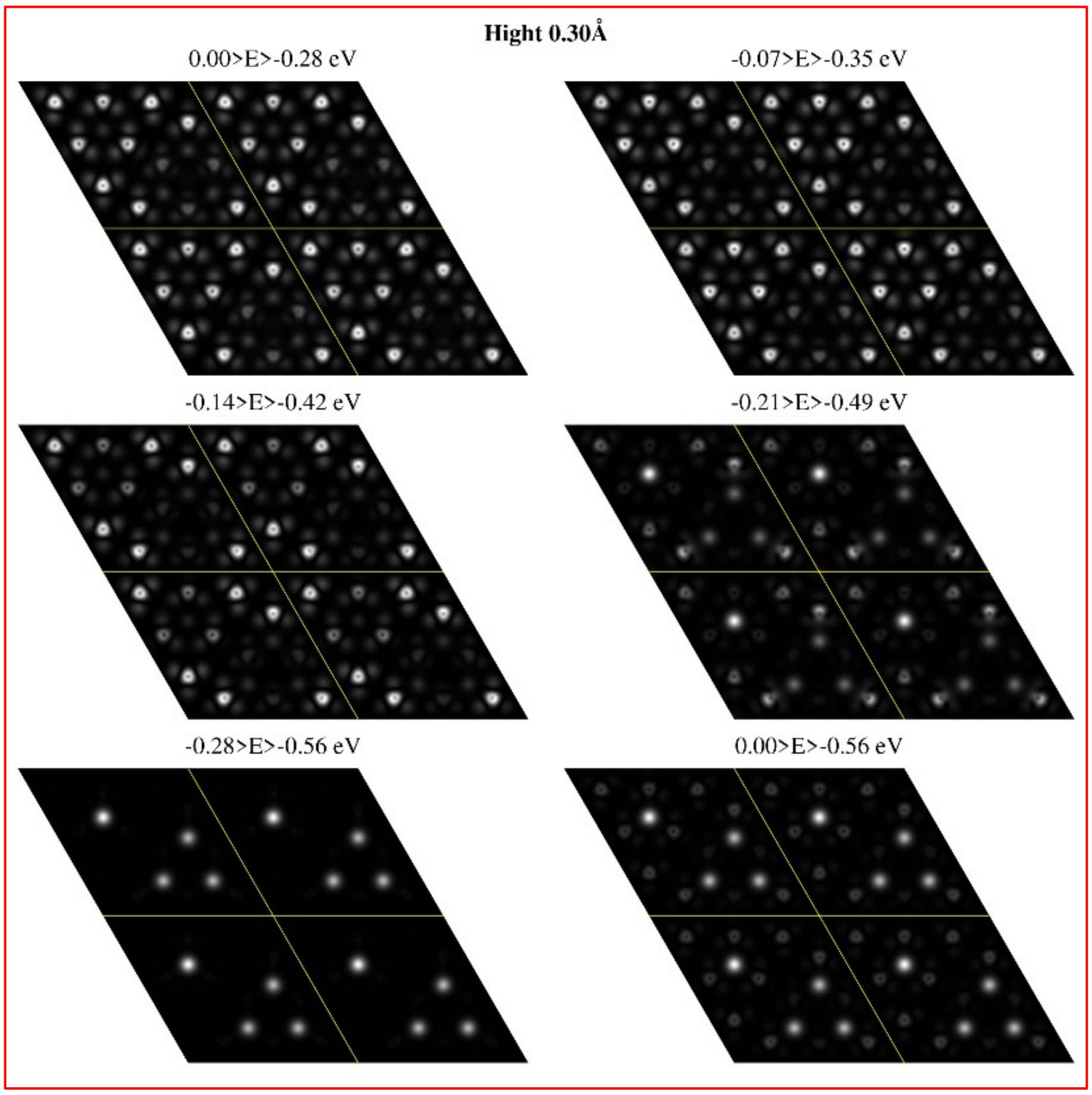

Fig.S16 STM of AA-d10-T12 at the hight of 0.3 Å

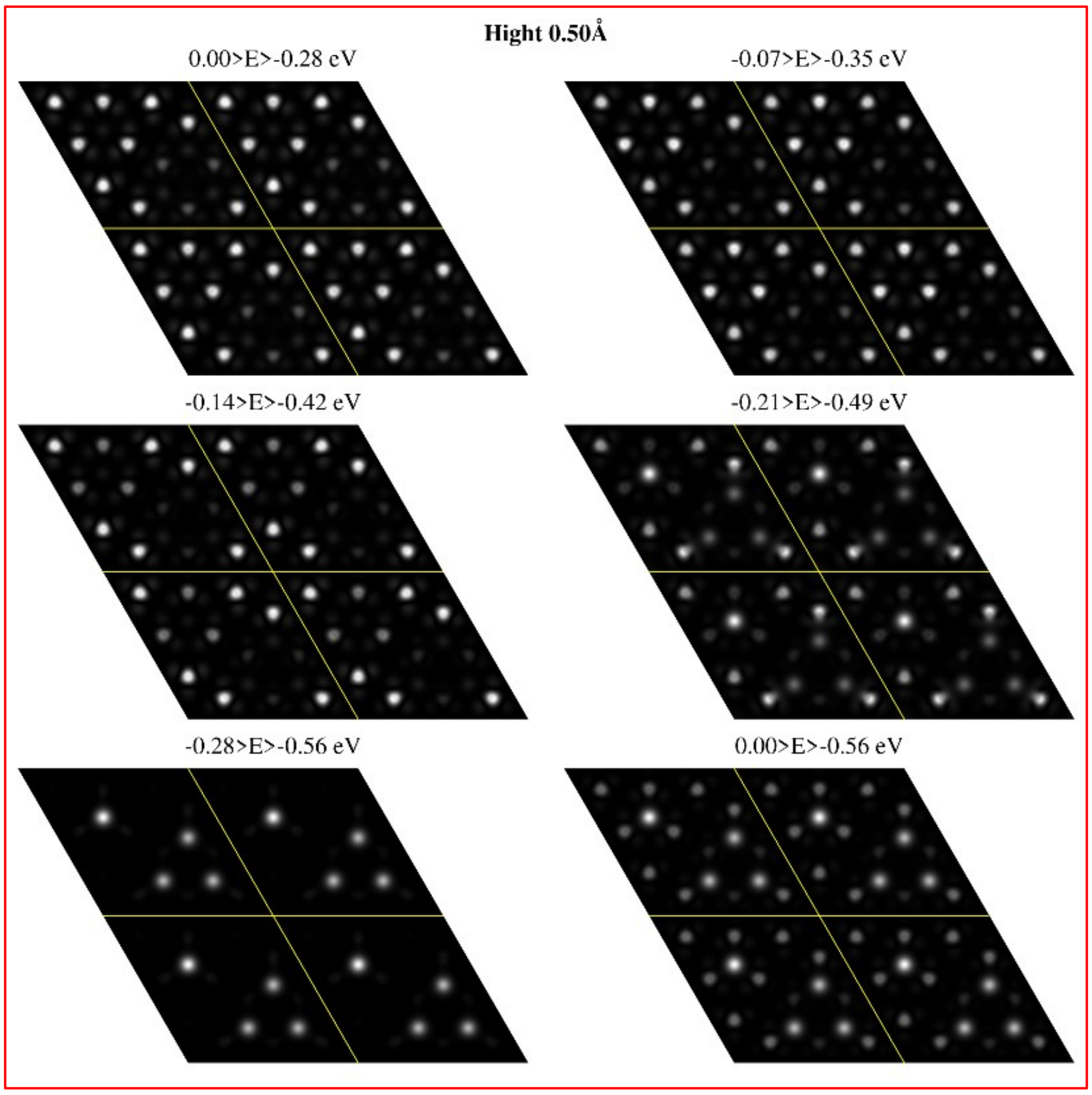

Fig.S17 STM of AA-d10-T12 at the hight of 0.5 Å

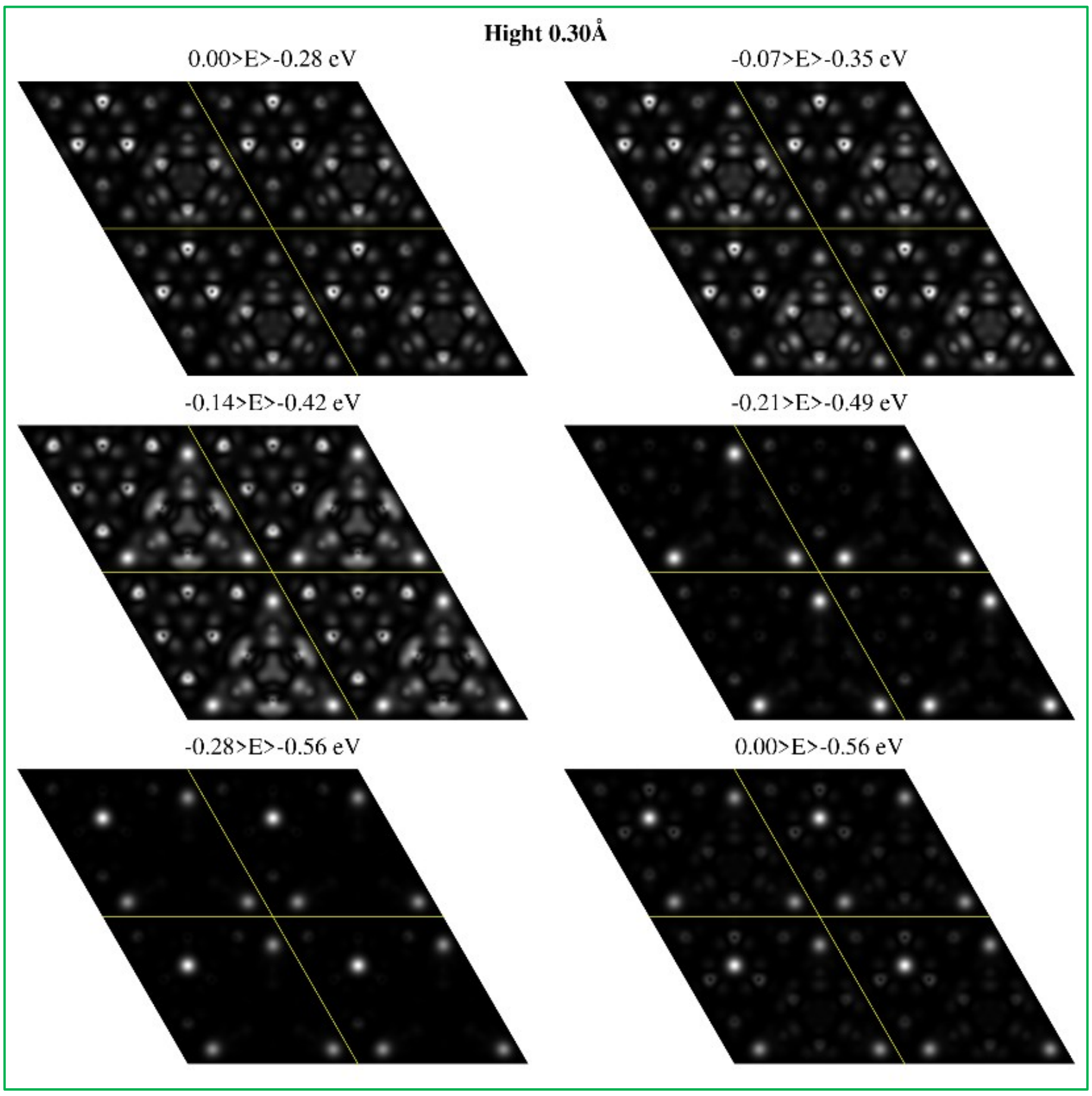

Fig.S18 STM of AA-d10-T9H3 at the hight of 0.3 Å

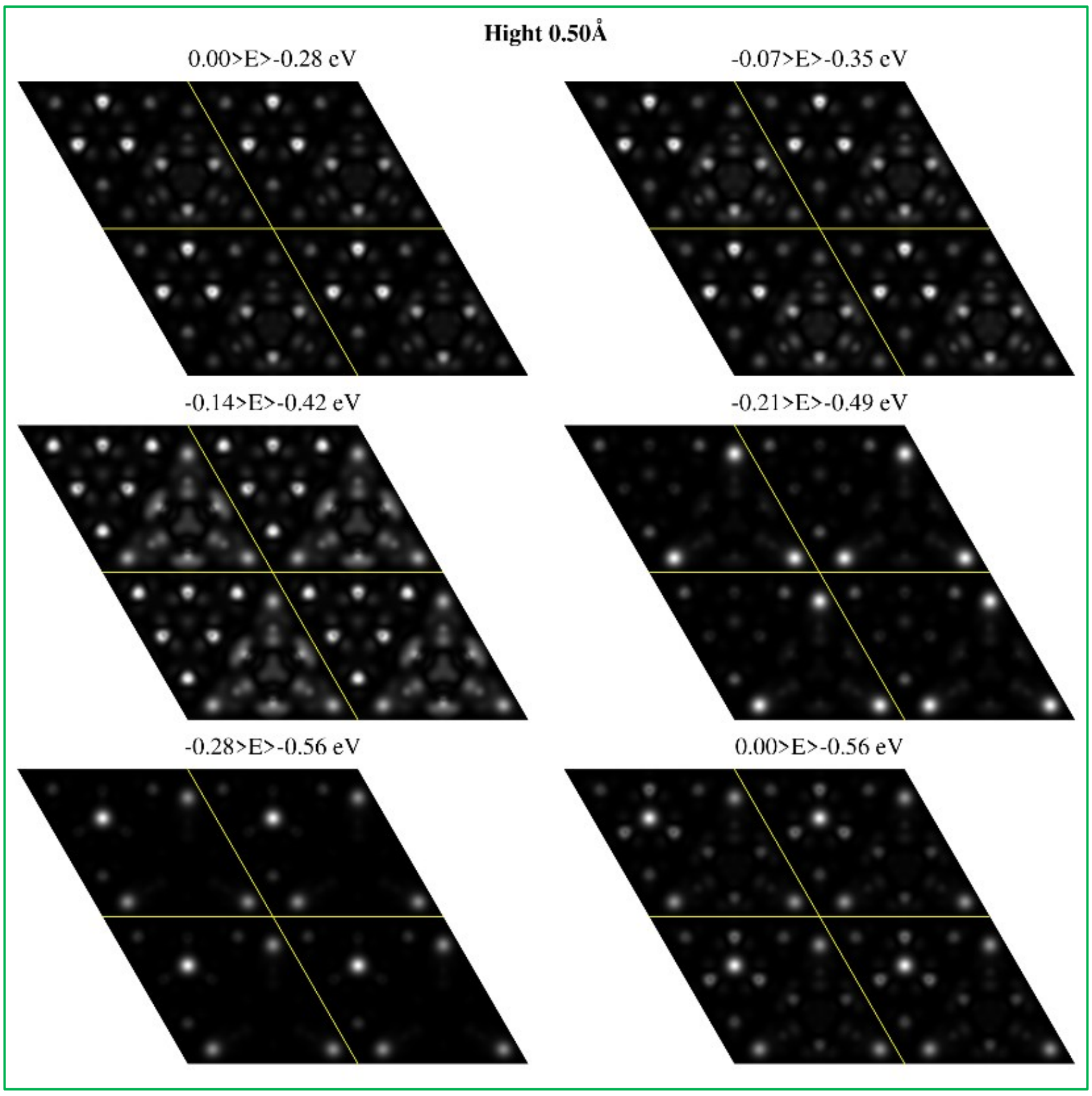

Fig.S19 STM of AA-d10-T9H3 at the hight of 0.5 Å